\pdfoutput=1

\documentclass[%
 reprint, longbibliography,
 amsmath,amssymb,
 aps,
]{revtex4-2}
\usepackage{graphicx} %
\usepackage{bm}
\usepackage[exponent-product = \cdot]{siunitx}
\DeclareSIUnit{\calorie}{cal}
\DeclareSIUnit{\Calorie}{\kilo\calorie}
\DeclareSIUnit{\atmass}{amu}
\usepackage{hyperref}
\hypersetup{colorlinks=true,linkcolor=blue,citecolor=blue,urlcolor=blue}

\usepackage{color}
\newcommand{\red}{}
\newcommand{\rred}{}

\usepackage{braket} %

\newcommand{\eu}{\mathrm{e}^}

\newcommand{\rmd}{\mathrm{d}}

\providecommand{\mat}[1]{\mathsf{#1}}
\renewcommand{\mathbf}[1]{\bm{#1}}

\newcommand{\eqn}[1]{Eq.\,(\ref{#1})}

\newcommand{\eqs}[1]{Eqs.\,(\ref{#1})}
\newcommand{\fig}[1]{Fig.~\ref{fig:#1}}

\newcommand{\secref}[1]{Sec.~\ref{sec:#1}\@}
\newcommand{\appref}[1]{Appendix~\ref{app:#1}\@}
\newcommand{\Refx}[1]{Ref.~\onlinecite{#1}}
\newcommand{\Refs}[1]{Refs.~\onlinecite{#1}} %

\usepackage{dcolumn} %
\newcolumntype{d}[1]{D{.}{.}{#1}} %

\usepackage{multirow}
\usepackage{placeins}
\usepackage{booktabs}

\begin{document}

\title{Evaluation of transition rates from nonequilibrium instantons}
\author{Eric R. Heller}
\email{hellere@berkeley.edu}
\author{David T. Limmer}
\email{dlimmer@berkeley.edu}
\affiliation{Department of Chemistry, University of California, Berkeley, 94720 Berkeley, USA}
\date{\today}

\begin{abstract}
Equilibrium rate theories play a crucial role in understanding rare, reactive events. However, they are inapplicable to a range of irreversible processes in systems driven far from thermodynamic equilibrium like active and biological matter. 
Here, we develop %
an %
\red{efficient numerical method to compute the rate constant of rare nonequilibrium events %
in the weak-noise limit based on an instanton approximation to the stochastic path integral and illustrate its wide range of application.} %
We demonstrate excellent agreement of the instanton rates with numerically exact results for a particle under a non-conservative force. We also study phase transitions in an active field theory. We elucidate how activity alters the stability of the two phases and their rates of interconversion in a manner that can be well described by modifying classical nucleation theory.

\end{abstract}

\maketitle

\section{Introduction}

Unlike their equilibrium counterparts, rare nonequilibrium transitions are not governed by the energetics of static transition states.
This renders efficient equilibrium techniques for computing rates and elucidating mechanisms largely invalid. 
Instead, reaction rates  in systems driven far from thermodynamic equilibrium are dependent on the details of their trajectories. This is formalized by Freidlin--Wentzell theory in the weak-noise limit with the optimal transition path or \textit{instanton}  \cite{Freidlin2012,Onsager1953,Graham1973}. 
Instantons have recently emerged as a powerful computational tool, offering insights into a wide variety of nonequilibrium phenomena across diverse disciplines like fluid dynamics, soft active matter, biology, and chemistry \cite{Weinan2004,Heymann2008,Heymann2008a,Kohn2006,Grafke2015,Grafke2017,Fuchs2022,ChavesOFlynn2011,Cruz2018,Zakine2023,Woillez2019,Bouchet2019,Kikuchi2020}. 

\red{In this work, we develop a %
numerically efficient instanton method to compute the rate constant of rare nonequilibrium transitions, which, in addition to the path action, includes a prefactor encoding fluctuations around the optimal path. This nonequilibrium instanton rate theory (NEQI) is derived through a rigorous asymptotic approximation of the exact path-integral representation of the transition rate and provides a quantitative understanding of transitions in complex driven systems.} 
We show the instanton rates computed with our method are in excellent agreement with numerically exact results at weak noise for a particle under a non-gradient force and illustrate dramatic changes of transition rates and mechanisms due to nonequilibrium driving. Further, we apply our method to an active field theory relevant to the study of how phase transforms are altered in the presence of active forces, revealing how insights from classical nucleation theory can be recovered even in nonequilibrium phase transitions \cite{cates2023classical,richard2016nucleation,redner2016classical}.

\section{Derivation of the nonequilibrium instanton rate equation}
\label{sec:derivation}

We consider a system with position vector ${\mat{x} = \{x_1,...,x_f\}}$ described by overdamped Brownian motion
\begin{align}
    \label{equ:langevin}
    \dot{\mat{x}}(t) = \mathbf{\mu} \mat{F}[\mat{x}(t)] +  \sqrt{2 \epsilon} \, \mathbf{\Lambda} \,\eta(t),
\end{align}
where $\dot{\mat{x}}(t) = \rmd \mat{x}/\rmd t$, $\mat{F}[\mat{x}(t)]$ is the sum of conservative and non-conservative forces exerted on the system, $\mathbf{D} = \mathbf{\Lambda}\mathbf{\Lambda}^{\!\text{T}}$ is the diffusion matrix, and $\mathbf{\mu} = \beta \epsilon \mathbf{D}$ is the mobility matrix at inverse temperature $\beta =1/k_\text{B} T$. 
The stochastic white noise, $\eta(t)$, defined via $\braket{\eta_i(t)} = 0$ and $\braket{\eta_i(t) \eta_j(t')} = \delta_{ij} \delta(t-t')$,
where $i,j \in [1,f]$, has a scale denoted by $\epsilon$.
\rred{While here we restrict ourselves to invertible diffusion matrices, we show in \appref{ud} how our method can be extended to systems with singular diffusion matrices such as those described by underdamped Langevin dynamics.}
We examine rate problems, where, in the absence of the nonequilibrium driving, an unstable transition state (TS) at $\mat{x}^\ddagger$ lies in between the metastable reactant and product configurations at $\mat{x}_\text{R}$ and $\mat{x}_\text{P}$.
All three configurations are fixed points of the force, but whereas the derivative matrix $\nabla F$ is negative definite at the reactant and product configurations, it exhibits one positive eigenvalue at the TS. This corresponds to the typical crossing of a potential-energy barrier, where an approximation to the rate is given by Kramers--Langer theory (KLT) \cite{Haenggi1990,Kramers1940,Langer1969,Langer1973,Pollak2023,Berezhkovskii2004}.

For systems driven far out of equilibrium by a non-conservative force, the steady-state probability distribution is generally not known, and KLT does not apply.
Nevertheless, the exact rate constant for a transition from reactant to product can be defined via \cite{Chandler1978TST} %
\begin{align}
    \label{equ:def_rate}
    k &= \frac{\braket{h_\text{R}(0) \dot{h}_\text{P}(t_\text{f})}}{\braket{h_\text{R}}}, %
\end{align}
where the brackets indicate an average over trajectories of duration $t_\text{rel} \ll t_\text{f} \ll 1/k$, and we assume the typical separation between the timescale of the system's local relaxation dynamics $t_\text{rel}$ and the timescale of the rare event $1/k$. 
The indicator functions $h_\text{R} = \Theta[\sigma^\ddagger-\sigma(\mat{x})]$ and $h_\text{P} =1-h_\text{R} $ define the reactant and product regions, which are separated by a dividing surface $\sigma^\ddagger$ along an order parameter $\sigma(\mat{x})$.
The flux--side correlation function can be written as an integral over the dividing surface parameterized by the $(f-1)$-dimensional vector $\mat{s}$
\begin{align}
    \label{equ:cfs}
    \braket{h_\text{R}(0) \dot{h}_\text{P}(t_\text{f})} &= \int \rho(\mat{x}_\sigma,t_\text{f} ; \mat{x}_\text{R},0) \, \dot{{x}}_\sigma(t_\text{f})  \, \rmd \mat{s}  ,
\end{align}
where $\rho(\mat{x}_\sigma,t_\text{f} ; \mat{x}_\text{R},0)$ is the joint probability density of starting at position $\mat{x}_\text{R}$ at $t=0$ and reaching the point $\mat{x}_\sigma$ on the dividing surface at time $t_\text{f}$, and $\dot{x}_\sigma=\nabla \sigma(\mat{x}) \cdot \dot{\mat{x}} $ is the velocity normal to the dividing surface. %
Because of the separation of timescales, the rate does not depend on the precise initial conditions as long as the probability density is concentrated within the reactant basin, which allowed the convenient choice $\rho(t=0) = \delta[\mat{x}(0) - \mat{x}_\text{R}]$ \cite{Lehmann2000,Lehmann2000a,Lehmann2003,Getfert2010}.

The probability density in \eqn{equ:cfs} can be expressed as an integral over all paths connecting $\mat{x}_\text{i}$ and $\mat{x}_\text{f}$ in time $t_\text{f}$,
\begin{align}
    \label{equ:rhoPI}
    \rho(\mat{x}_\text{f},t_\text{f} ; \mat{x}_\text{i},0) &= \int_{\mat{x}(0)=\mat{x}_\text{i}}^{\mat{x}(t_\text{f})=\mat{x}_\text{f}} \eu{-S[\mat{x}(t)] / \epsilon} \, \mathcal{D} [\mat{x}(t)] ,
\end{align}
each of which is exponentially weighted by its Onsager--Machlup
action $S[\mat{x}] = \int_{0}^{t_\text{f}} \mathcal{L}(\mat{x},\dot{\mat{x}}) \, \rmd t$ \cite{Onsager1953,Graham1973,Freidlin2012}, where the time-dependence of positions and velocities is implied, and the Lagrangian is given by 
\begin{align}
    \label{equ:Lagrangian}
    \mathcal{L}(\mat{x},\dot{\mat{x}}) &= \frac{1}{4}[\dot{\mat{x}} - \mathbf{\mu}\mat{F}(\mat{x})]^\text{T}  \mathbf{D}^{-1} \,[\dot{\mat{x}} - \mathbf{\mu}\mat{F}(\mat{x})] .
\end{align}
The exact evaluation of \eqn{equ:rhoPI} is unfeasible for complex systems because it involves a sum over infinitely many paths. Some importance sampling methods exist to estimate it for systems away from equilibrium \cite{VPS,Allen2009}.
In the weak-noise limit, however, we can use Laplace's method to devise an efficient approximation about the minimum-action path or instanton, $\bar{\mat{x}}(t)$.

The instanton, which can be interpreted as the optimal transition path, constitutes the centerpiece of our rate \red{method}. The action of the instanton is determined by its endpoints and the propagation time so that it can be written as $S \equiv S(\mat{x}_\text{i},\mat{x}_\text{f};t_\text{f}) = \int_{0}^{t_\text{f}} \bar{\mat{p}}^\text{T}\! \mathbf{D} \bar{\mat{p}} \, \rmd t$, 
where
$\bar{\mat{p}}= \mathbf{D}^{-1} [\dot{\bar{\mat{x}}} - \mathbf{\mu}\mat{F}(\bar{\mat{x}})]/2$ is the canonical momentum along the instanton.
The Lagrangian in \eqn{equ:Lagrangian} can be Legendre transformed to obtain the corresponding Hamiltonian, whose value $E = \bar{\mat{p}}^\text{T} \mathbf{D} \bar{\mat{p}} + \bar{\mat{p}}^\text{T} \mathbf{\mu} \mat{F}(\bar{\mat{x}})$ is conserved along the instanton because we consider forces without explicit time dependence. 
The Hamilton--Jacobi formalism naturally yields the well-known relations for the instanton energy and the momentum at the endpoint \cite{LandauMechanics,GoldenInverted}  
\begin{align}
    \label{equ:momentum}
    \bar{\mat{p}}_\text{f} &\equiv \bar{\mat{p}}(t_\text{f}) = \frac{\partial S}{\partial \mat{x}_\text{f}},
    \qquad
    \frac{\partial S}{\partial t_\text{f}} = -E .
\end{align}
For the Laplace approximation of the rate, we require an instanton from %
reactant to product that minimizes the action not only in coordinate space but also in time $t_\text{f}$, which is achieved when $E = 0$ or equivalently $t_\text{f}\rightarrow\infty$. 
Consequently, the instanton takes the form of a ``double kink'' illustrated in \appref{timetranslation}, connecting the three fixed points of the force, even in the presence of a non-gradient force. 
In spite of the infinite propagation time, the action remains finite because the dwell phases do not contribute. 

From the stationarity conditions $\delta S = 0$ and $E=0$ follows  $\dot{\bar{\mat{x}}}^2 = [\mathbf{\mu}\mat{F}(\bar{\mat{x}})]^2$.  
At equilibrium, the action is directly related to the height of the potential-energy barrier because the instanton minimizes the potential along its path, thus aligning with the mobility-scaled conservative force $\mathbf{\mu}\mat{F}_\text{c} = -\mathbf{\mu} \nabla V$.
Consequently, $\dot{\bar{\mat{x}}} = \pm\mathbf{\mu}\mat{F}_\text{c}$, where the minus sign corresponds to the %
\textit{activation path} from reactant to TS\@, and the plus sign corresponds to the \textit{relaxation path}, which follows the force into the product configuration.
Notably, only the activation path contributes to the action.
Unlike the relaxation path, which remains aligned with the force also under nonequilibrium conditions, the activation path generally deviates, leading to changes in transition rate and mechanism.

Since the relaxation path does not contribute to the rate \cite{Luckock1990}, the natural choice for the location of the dividing surface is the unstable fixed point of the total force, to which we will refer as a TS\@. Therefore, we can focus solely on the activation path, which we henceforth call the instanton.
Combining \eqs{equ:def_rate}, \eqref{equ:cfs} and \eqref{equ:rhoPI} leads to the path-integral representation of the rate constant
\begin{align}
    \label{equ:PI_laplace}
    k \, Z_\text{R} =
    \lim_{N\rightarrow \infty}  \int \frac{\dot{x}_\sigma(t_\text{f}) \, \eu{-S_N/\epsilon} \, \rmd \mat{x}_1... \rmd \mat{x}_{N-1} \rmd \mat{s}}{(\text{det}[4\pi\epsilon \tau \mathbf{D}])^{N/2}} ,   
\end{align}
where, in order to evaluate the action numerically, we discretized the paths into $N$ equal-time segments of length $\tau = t_\text{f}/N$ \cite{Pirey2022}.  In the instanton limit, the reactant path partition function $Z_\text{R}$=1.
The discretized action $S_N\equiv S_N(\mat{x}_0,..., \mat{x}_{N};t_\text{f})$ is a function of the ``beads'' $\mat{x}_n$, which constitute copies of the system along the path
\begin{multline}
    \label{equ:SN}
    S_N = \frac{\tau}{4} \sum_{n=0}^{N-1} \left[\frac{\mat{x}_{n+1} - \mat{x}_n}{\tau} - \mathbf{\mu} F(\mat{x}_n)\right]^\text{T}\\  
    \times \mathbf{D}^{-1} \left[\frac{\mat{x}_{n+1} - \mat{x}_n}{\tau} - \mathbf{\mu} F(\mat{x}_n)\right] .
\end{multline}
We aim to evaluate the integrals over the beads and the dividing surface with Laplace's method.
Hence, the instanton with endpoints $\mat{x}_0=\mat{x}_\text{R}$ and $\mat{x}_N=\mat{x}^\ddagger$ must satisfy $\nabla_\text{s} S_N = 0$, where $\nabla_\text{s} = (\partial/\partial \mat{x}_1,..., \partial/\partial \mat{x}_{N-1}, \partial/\partial \mat{s})$ is the gradient with respect to the positions of all intermediate beads and the parameters of the dividing surface.
The optimal path can therefore be located with standard optimization algorithms \cite{Fletcher,Weinan2004}
(see \appref{optimization}). 
From the stationarity condition $\partial S_N/\partial s_i = \bar{\mat{p}}_\text{f} \cdot \partial \mat{x}_N/\partial s_i = 0$, we find that the momentum at the endpoint is orthogonal to the dividing surface \cite{Lehmann2003}. %
Given that the mobility-scaled force and the instanton velocity are parallel at the TS\@, it follows that $\dot{\bar{x}}_\sigma(t_\text{f}) = \bar{\mat{p}}_\text{f}^\text{T} \mathbf{D} \bar{\mat{p}}_\text{f} / \bar{p}_\text{f}$, where $\bar{p}_\text{f} = \lVert \bar{\mat{p}}_\text{f} \rVert$.

\red{Note that the second-derivative matrix of the action, $\nabla_\text{s}^2 S_N$, remains symmetric even for non-conservative forces, where $\nabla F$ is not symmetric.}
Approximating \eqn{equ:PI_laplace} by Laplace's method 
now appears straightforward. 
However, the instanton, as an infinite-time trajectory, %
can dwell at the reactant or the TS for an arbitrary time without changing its action (see \appref{timetranslation}). Hence, there is an infinite number of instantons that are related by continuous time translation. This leads to the emergence of a Goldstone mode, which manifests itself as a zero eigenvalue in %
$\nabla_\text{s}^2 S_N$, and thus cannot be integrated over by Laplace's method.
Previous work avoided this difficulty by considering explicitly time-dependent forces that break time-translation symmetry \cite{Lehmann2000,Lehmann2000a,Lehmann2003}.

\red{However, progress can be made by separating the integral over the Goldstone mode from the integral over the other, orthogonal modes \cite{Faddeev1967}} and noting that a small translation along the zero mode can be related to a time translation of the instanton, $\rmd \zeta_0 = \sqrt{B_t/\tau} \, \rmd t'$ \cite{InstReview,Luckock1990},
where $B_t = \int \dot{\bar{\mat{x}}}^2 \,\rmd t$ %
is a normalization constant \footnote{Note that $B_t = S$ for an instanton of a system at equilibrium with isotropic diffusion, where $\mathbf{D}=\mathbb{I}$.}.
This allows us to transform the integral over the Goldstone mode to an integral over a collective time, e.g., the time to reach the TS\@. %
By noting that the prefactor $\bar{\mat{p}}_\text{f}^\text{T} \mathbf{D} \bar{\mat{p}}_\text{f}$ is identical to the instanton energy [\eqn{equ:momentum}], %
this time integral is readily solved by substitution, leading to the nonequilibrium instanton formula for the rate constant  
\begin{align}
    \label{equ:k_neqi}
    k &\sim \frac{1}{4\pi \tau}
    \sqrt{\frac{B_t}{\tau \bar{p}_\text{f}^{2} \Sigma}}  \, \eu{-S_N/\epsilon},
\end{align}
where $\Sigma = \det'[2\tau \nabla^2_\text{s} S_N] (\text{det}[\mathbf{D}])^{N}$ encodes fluctuations around the optimal path and the prime indicates omission of the zero eigenvalue. Note that because $\Sigma \propto \bar{p}_\text{f}^{\,-2}$ \cite{Lehmann2000,Lehmann2003}, the term $\bar{p}_\text{f}^2\Sigma$ can be treated as a slowly varying prefactor despite $\bar{p}_\text{f}$ going to zero in the infinite-time limit \cite{BenderBook,GoldenCI}.
All quantities in \eqn{equ:k_neqi} are efficiently evaluated from a single path, the instanton. 
By construction, \eqn{equ:k_neqi} becomes exact in the limit $\epsilon \rightarrow 0$ and reduces to KLT at equilibrium (see \appref{kramers} for details) \footnote{If a system has several instantons (e.g., due to symmetry) that are sufficiently far separated in path space, the rate is given by the sum over the individual contributions.}.
\red{While we concentrated on the case of additive Gaussian noise throughout our derivation, we show in \appref{multiplicative} that the rate equation \eqn{equ:k_neqi} extends straightforwardly to the case of multiplicative Gaussian noise, where $\mathbf{\Lambda}\rightarrow\mathbf{\Lambda}(\mat{x})$. %

It is worth noting that the coupling of strongly anharmonic modes to the reaction coordinate leads to inaccuracies of the asymptotic approximation, as is common for instanton approaches \cite{InstReview}. 
In extreme cases, where the action landscape becomes almost flat in a certain region, higher-order terms in $\epsilon$ become significant and may even alter the transition mechanism \cite{Maier1992,Boerner2023}. In order to obtain an accurate rate estimate in these cases, higher-order terms must be included \cite{Lawrence2023,Ekstedt2022} or sampling approaches need to be employed \cite{VPS,Allen2009}. 
}

Since the instanton rate is defined in the infinite-time limit, in practice, the propagation time is extended until convergence is reached, while ensuring convergence also with respect to the number of beads.
\red{The rate of convergence may be enhanced by allowing for variable time intervals in order to avoid clustering of beads near the fixed points of the force, as shown in \appref{timestep} \cite{Zhou2008,Rommel2011grids,Cvitas2016instanton}.}
We can leverage the banded structure of $\nabla^2_\text{s} S$ to reduce memory requirements and the scaling of the diagonalization to $\mathcal{O}(N^2)$. \red{The results presented in the following are generated with \eqn{equ:k_neqi}.
In \appref{trajectory}, we derive an equivalent instanton rate formula in the ``trajectory formalism'' \cite{Hunt1981}, which is numerically less robust but further reduces the scaling with respect to the number of beads to $\mathcal{O}(N)$ \cite{Althorpe2011ImF,DaC,GoldenRPI,InstReview,GoldenGreens,GoldenInverted}. Similar to \Refx{Bouchet2016}, the equation also lends itself to solutions through matrix differential equations and related techniques \cite{Schorlepp2021,Grafke2023,Schorlepp2023,Bouchet2022,Lehmann2003,Maier1997,Schorlepp2023a}, and the numerically most efficient approach will generally depend on the system under consideration.}

\section{Driven particle model}
\begin{figure}[b]
    \centering
    \includegraphics[width=0.48\textwidth]{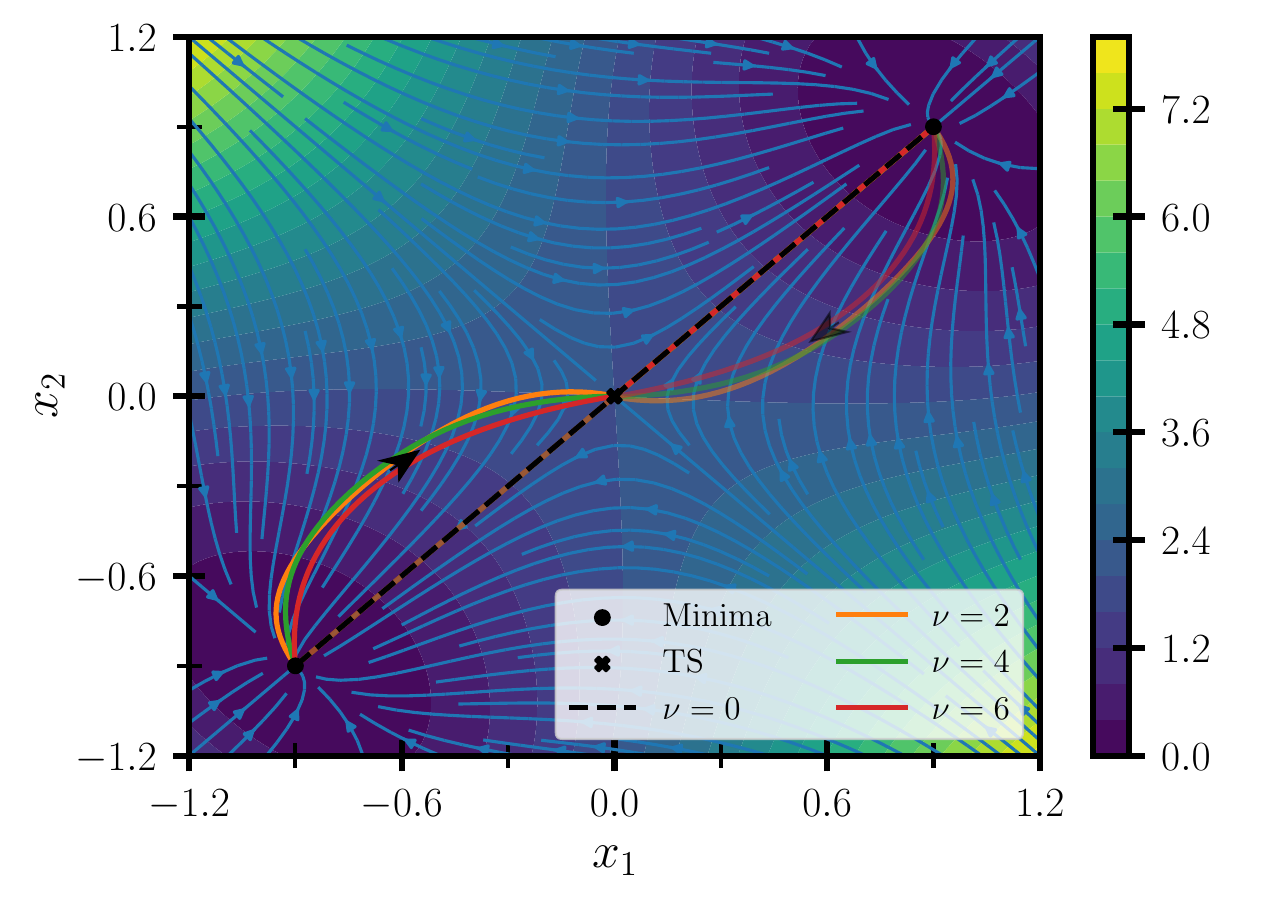}
    \caption{Instantons at different values of $\nu$ on a color map of the potential defined in the text and a stream plot depicting the conservative force field. 
    The forward (solid) and backward (transparent) instantons, whose direction is indicated by the black arrows, do not coincide out of equilibrium ($\nu\neq 0$).%
    }
    \label{fig:pathways}
\end{figure}
We apply our new method to a model system in two dimensions with total force
\begin{align}
    \label{equ:force}
    \mat{F}(\mat{x}) & = -\frac{\partial V(\mat{x})}{\partial \mat{x}} + \nu    
    \begin{pmatrix}
    x_{-}^3 - x_{-} \\
    x_{-}^3 - x_{-} 
    \end{pmatrix} ,
\end{align}
where $x_{\pm} = x_1 \pm x_2$, and the non-conservative force is scaled by the driving strength $\nu$. We consider the potential $V(\mat{x}) = A(x_{+}^2-B^2)^2/B^4 + x_{-}^2$ with $A=2$ and $B=1.8$, and set $\mathbf{\mu} = \mathbf{D} = \mathbb{I}$ for simplicity, where $\mathbb{I}$ is the identity matrix.
In Fig.~\ref{fig:pathways}, we illustrate how nonequilibrium driving alters the transition mechanism. At equilibrium, the instanton follows the minimum-potential pathway along $x_{+}$, whereas the activation paths in the driven systems are curved, thus achieving a lower action. %
It can be seen that the forward and backward instantons lie on top of one another at equilibrium, in agreement with semiclassical instanton theories based on linear response \cite{InstReview,PhilTransA,MLJ}.
Conversely, the forward and backward transition mechanisms generally differ for driven processes ($\nu>0$) because they break detailed balance. 
 \begin{figure}
    \centering
    \includegraphics[width=0.48\textwidth]{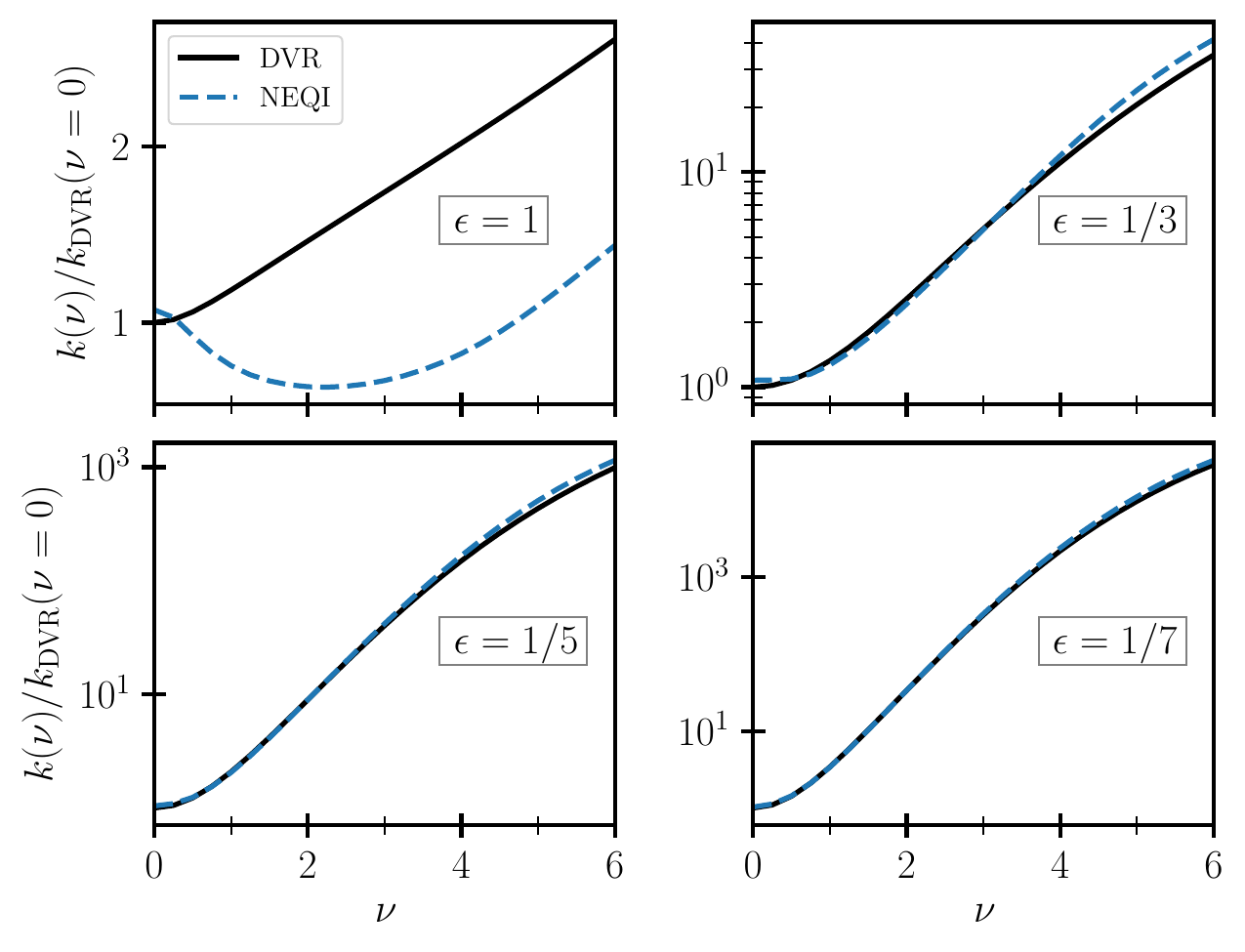}
    \caption{Rate constants for the model defined in \eqn{equ:force} over a range of noise strengths $\epsilon$ and driving strengths $\nu$ computed from nonequilibrium instanton theory (NEQI) and a numerically exact discrete variable representation (DVR) of the Fokker--Planck operator. For each $\epsilon$, the results are given relative to the DVR equilibrium rate ($\nu = 0$).
    }
    \label{fig:rates}
\end{figure} 

We present the rate constants computed from these instantons for various values of the noise strength in Fig.~\ref{fig:rates} alongside numerically exact results from a discrete variable representation of the Fokker--Planck operator \cite{Colbert1992DVR,Piserchia2015,Shizgal2015}. The numerical details of the rate calculations are reported in \appref{computational}. Generically, the transition rate increases with the addition of the nonequilibrium driving, consistent with expectations from thermodynamic speed limits \cite{kuznets2021dissipation}. Our NEQI method captures the dramatic speed ups of the rate due to nonequilibrium driving and converges to the exact result at weak noise as expected.
\begin{figure*}[t]
    \centering
    \includegraphics[width=0.9\textwidth]{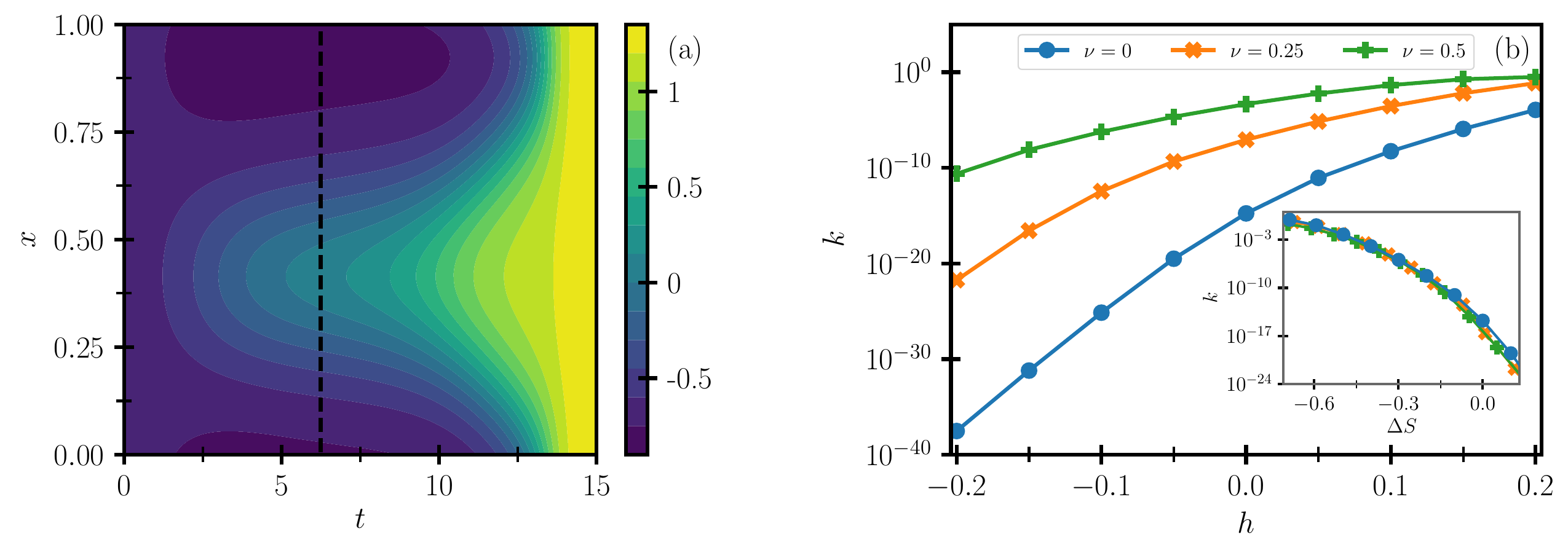}
    \caption{Instanton results for the active field theory defined in the text with parameters $\kappa = 0.01$, $u = 1/4$, $a = -1/2$, $\mu=1$. (a) Instanton at $h=0.1$ and $\nu=0.5$, where the color map shows the configuration of the field $\phi$ as a function of space and time along the path. 
    The phase transition proceeds from the uniform reactant state ($\phi\approx -0.71$) to the uniform product state ($\phi\approx 1.32$) via a non-uniform TS\@, whose location is indicated by the dashed line.
    At the TS\@, part of the field extends towards the product, forming a critical nucleus.
    (b) Instanton rate constants over a range of applied fields $h$ and driving strengths $\nu$ with $\epsilon=5\cdot 10^{-3}$.
    The inset shows the same rate constants as a function of the difference between the instanton actions of the forward and the backward processes, $\Delta S$.
    }
    \label{fig:nucleation}
\end{figure*}

\section{Nucleation in an active field theory}
\label{sec:activefield}

To illustrate the generality of our approach, we next consider the phase transformation of a collection of active particles. Rather than consider an explicit particle model, we employ a coarse-grained representation or active field theory known as Active model A dynamics \cite{Cates2015,Caballero2020}, which has proven fruitful in understanding the emergent phenomena of these systems on large length scales. 
The Langevin equation for a non-conserved scalar field $\phi$ in $f$ dimensions is
\begin{align}
    \dot{\phi}(\mat{x},t) &= \mu F[\phi(\mat{x},t)] + \sqrt{2\epsilon} \, \eta(\mat{x},t) , 
\end{align}
where the mobility $\mu$ is constant, the white noise is defined via $\braket{\eta(\mat{x},t) \eta(\mat{x}',t')} = \delta(\mat{x}-\mat{x}')\delta(t-t')$, and  $F(\phi) = - \delta V/\delta \phi + F_\text{nc}(\phi)$ is the drift that again contains gradient and non-gradient terms.
Our nonequilibrium instanton theory directly extends to such field-theoretical descriptions, where the action,
\begin{align}
    \label{equ:fieldaction}
    S[\phi(\mat{x},t)] &= \frac{1}{4} \int_0^{t_\text{f}} \int \left( \dot{\phi}(\mat{x},t) - \mu F[\phi(\mat{x},t)] \right )^2 \rmd \mat{x} \, \rmd t ,
\end{align}
must be discretized not only in time but also in space. We take the potential to be $V[\phi] = \int [\kappa \, |\nabla \phi |^2/2 + a \phi^2+u \phi^4   - h\phi  ] \rmd \mat{x}$, a sum of a square-gradient term with stiffness $\kappa$ and a free-energy density parameterized by $a<0$, $u>0$ and field $h$. The time-reversal symmetry of the system is broken by adding the strongly non-local, non-conservative force $F_\text{nc}[\phi(\mat{x})] = \nu \int \phi^2(\mat{x}') \, \rmd \mat{x}'$ \cite{Zakine2023}.

\red{It is well known that} an additional complication arises because transitions of fields typically involve the formation of an interface. 
A system with periodic boundary conditions is invariant under translation of the interface, leading to up to $f$ zero modes at the TS \cite{Langer1969}. However, similar to the time-translation mode, the integral over these zero modes can be transformed to an integral over space \cite{Cottingham1993}, where $B_x = \int \kappa |\nabla \phi^\ddagger|^2 \, \rmd \mat{x}$ is the Jacobian of the transformation, and $\phi^\ddagger(\mat{x})$ is the TS configuration of the field. 
Along with the volume factor $\mathcal{V}$ from the integral over space, this leads to the instanton expression for the nonequilibrium rate constant in field space
\begin{align}
    \label{equ:k_neqi_fieldT}
    \frac{k}{\mathcal{V}} &\sim \frac{1}{4\pi \tau}
    \sqrt{\frac{B_t}{\tau \xi^f \bar{p}_\text{f}^2 \Sigma}} \, \left(\frac{B_x}{4\pi\epsilon\tau\kappa}\right)^{\!f/2}   \eu{-S_N/\epsilon},
\end{align}
which reduces to the field-space generalization of KLT at equilibrium \cite{Langer1969,Ekstedt2022a,Cottingham1993,Simeone2023}.
Here, $\xi$ is the spatial resolution of the grid \footnote{For brevity we choose the same grid spacing $\xi$ along all degrees of freedom. Choosing different spacings corresponds to the replacement $\xi^f \rightarrow \prod_{j=1}^f \xi_j$.}, and $B_t = \int \dot{\bar{\phi}}(\mat{x},t)^2  \, \rmd \mat{x} \, \rmd t$.  As before, the instanton $\bar{\phi}(\mat{x},t)$ is the minimum-action path from reactant to TS\@, where each bead $\phi_n$ now corresponds to a field configuration. 
Hence, the instanton satisfies $\nabla_\phi S = 0$, where $\nabla_\phi$ is the derivative vector with respect to all intermediate beads and the parameters of the dividing surface. The fluctuation determinant $\Sigma = \text{det}'[2\tau  \nabla_\phi^2 S_N /\xi^f]$ has the modes from time translation and translation of the interface removed, 
and $\bar{p}_\text{f} = \partial S_N/ \partial \phi_N /\xi^f$ (see also \appref{activefield}).

We can now employ our nonequilibrium instanton theory to study nucleation in the active model A\@.  The field is defined in a one-dimensional, periodic box of length $1$ discretized by $65$ points.
An optimal transition path of this system starting in the $\phi<0$ phase and transitioning to the $\phi>0$ phase is illustrated in Fig.~\ref{fig:nucleation}(a). The path exhibits the formation of a critical nucleus at the TS\@, where the field partially extends from the reactant towards the product. In Fig.~\ref{fig:nucleation}(b), we present the instanton rate constants for this phase transition over a range of applied fields and driving strengths.
Notably, the driving accelerates the rates for all values of $h$, which is accompanied by a decrease in the size of the critical nucleus required for a transition (see \appref{activefield}).
The speed-up is most pronounced for processes with high equilibrium barriers.
Despite these nonequilibrium effects, the inset of Fig.~\ref{fig:nucleation}(b) reveals that the rate curves for different $\nu$ nearly collapse when one accounts for the relative stability of the two phases using the difference in instanton action between forward and backward processes.
This suggests that the activity primarily shifts the coexistence line between the two phases, not alters the effective surface tension. This shift could be incorporated into a modified classical nucleation theory through a renormalized chemical potential \cite{Cho2023}. Extending this analysis to conservative field theories like active model B would be an interesting future direction \cite{wittkowski2014scalar}.

\section{Conclusions}
\red{In conclusion, we developed an instanton method to compute the rate constant of rare nonequilibrium transitions, which is derived through a rigorous weak-noise approximation of the exact rate expression.  %
The numerical efficiency of the method allows applications to complex high-dimensional systems in either particle or field-theory representations. 
NEQI combines quantitative estimates of the nonequilibrium rate with the intuitive insight into the transition mechanism from the optimal path, which complements nonequilibrium sampling approaches also outside of the weak-noise limit \cite{VPS,Allen2009}.
Due to its generality, efficiency and conceptual simplicity, we believe the method to find wide application in the description of rare transitions in dissipative systems across scales.}\\

The source code for an instanton calculation and the data that supports the findings of this study are openly available online \cite{SourceData}.\\
\\
\section*{Acknowledgements}
ERH is grateful for financial support from the Swiss National Science Foundation through Grant 214242. DTL was supported by NSF Grant No.\ CHE-1954580 and the Alfred P. Sloan Foundation. 
The authors thank Dr.\ Jorge L.\ Rosa-Raíces for support with the DVR calculations.

\appendix

\section{Time-translation symmetry}
\label{app:timetranslation}

\begin{figure}[t]
    \centering
    \includegraphics[width=0.48\textwidth]{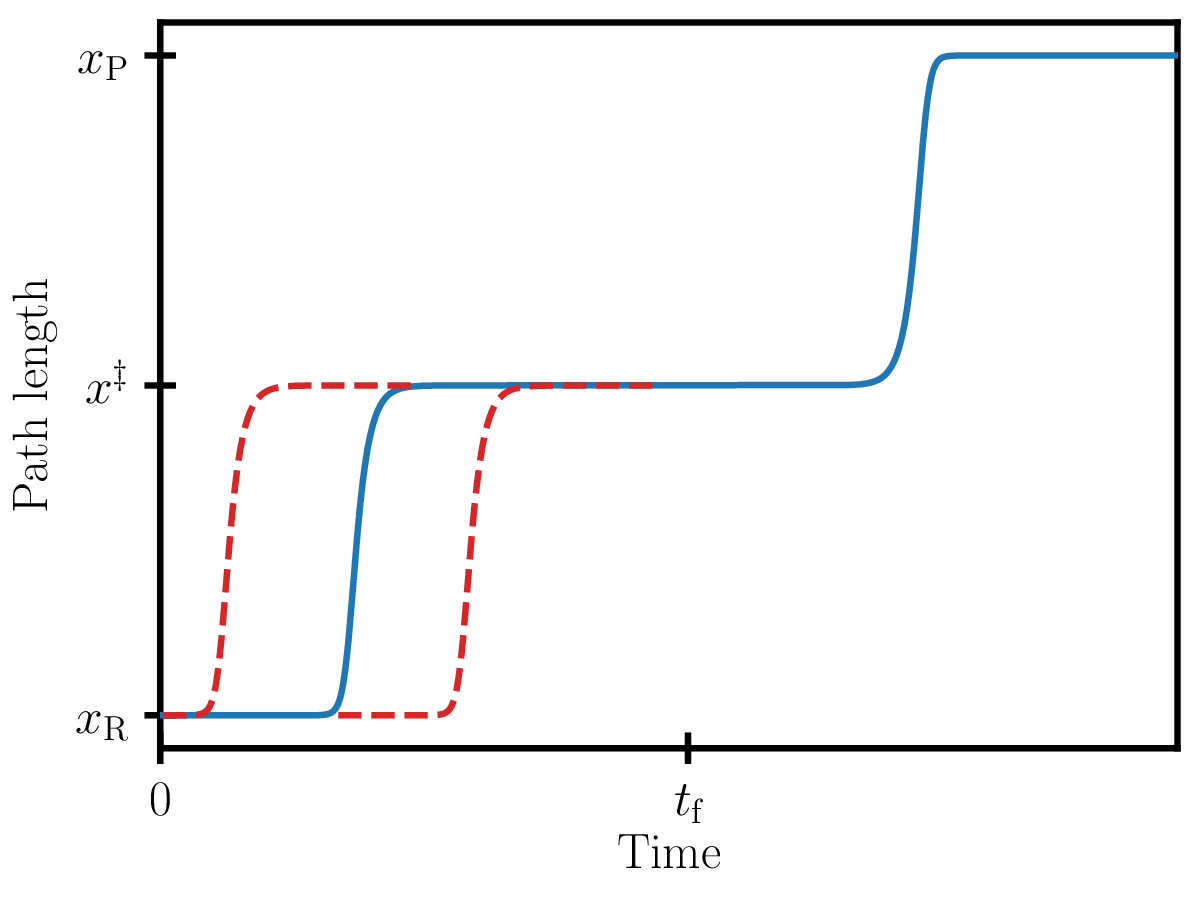}
    \caption{Plot of the path length as a function of time along an instanton in the long-time limit. The comparatively fast transitions between the fixed points of the force are separated by long dwell phases at the reactant, product and TS configurations, giving rise to the typical double kink shape. The activation path reaches the TS at time $t_\text{f}$. The red dashed paths indicate equivalent activation paths related through time translation.}
    \label{fig:doublekink}
\end{figure}

As described in \secref{derivation}, the instanton is an infinite-time trajectory. The typical ``double kink'' shape of an optimal transition path between reactant and product is illustrated in Fig.~\ref{fig:doublekink}, where the path length is shown as a function of time. 
It can be seen that the instanton may dwell for an arbitrary, quasi-infinite time at the reactant, product or TS configurations (the fixed points of the force) without changing its action because force, velocity and canonical momentum go to zero at these points [\eqn{equ:Lagrangian}]. Hence, there exists an infinite number of instantons related through time translation, as indicated in Fig.~\ref{fig:doublekink}.

The rate-determining activation path, connecting reactant and TS\@, corresponds to the first kink in Fig.~\ref{fig:doublekink}. 
In numerical calculations, the long-time limit is reached when the activation process is fully resolved, which is fast compared to the infinitely long dwell phases.
The time-translation symmetry of the activation path manifests itself as a zero eigenvalue in the  second-derivative matrix of the action, rendering a ``na\"ive'' Laplace approximation of the rate divergent.
We address this issue in \secref{derivation} by transforming the integral over the zero mode to a collective time integral, as encountered in the evaluation of flux--flux correlation functions, which must be solved separately.

\section{Instanton optimization}
\label{app:optimization}

Locating an instanton consists in the minimization of the discretized Onsager--Machlup action [\eqn{equ:SN}], where the endpoints are fixed to the reactant and TS configurations. The TS can be located either by optimizing a saddle point of the force directly or by locating the full minimum-action pathway between reactant and product.
A whole host of established minimization routines may be put to use for the instanton optimization. In this work, we employ conjugate-gradient and Newton--Raphson algorithms that minimize the magnitude of the gradient vector of the discretized action with respect to all intermediate bead positions.
For the latter, a small shift must be added to the diagonal of the second-derivative matrix to avoid problems with zero eigenvalues \cite{InstReview}. The banded, symmetric properties of the second-derivative matrix of the action may be leveraged to significantly improve the memory and time efficiency of Newton--Raphson optimizers for large systems. Furthermore, we update the second-derivative matrix during the optimization \cite{Fletcher}, requiring only a single calculation of the second-derivative matrix for the whole minimization.  
Alternatively, specialized instanton optimization routines have been suggested \cite{Heymann2008,Kikuchi2020,Zakine2023}. 

The instanton rate must be converged with respect to both the number of beads, $N$, and time, $t_\text{f}$.  For the latter, a plateauing plot of the path length, as in Fig.~\ref{fig:doublekink}, indicates that the long-time limit is reached. Additionally, the numerical value of the ``zero eigenvalue'' from time translation can be monitored because the time-translation symmetry is only approximate for long but finite time. As $t_\text{f}$ is increased, the ratio between the zero eigenvalue and the remaining eigenvalues decreases. A sufficiently small ratio thus indicates convergence. However, the most reliable metric for convergence remains the comparison between actions and rates at different values of $t_\text{f}$ and $N$. 
We have found that it is often useful to first converge a minimum-action path with a relatively short time based on a simple initial guess, e.g., equidistant beads along a straight line connecting the endpoints. The result can then be used as an initial guess for subsequent optimizations with gradually increasing $t_\text{f}$ and $N$.

\section{Computational details}
\label{app:computational}

The instanton rates for the particle model in \fig{rates} are calculated with $N=800$ beads and a propagation time of $t_\text{f}=4$, which corresponds to convergence to within less than $1\%$ error. An example of a convergence table is shown in Table~\ref{tab:convergence}. As expected, for longer propagation times, more beads are required to converge the rate. In order to judge whether the rate is converged, it is particularly useful to look at the ``diagonal convergence'' in Table~\ref{tab:convergence}, i.e., comparing rates computed at the same time step $\tau = t_\text{f}/N$. Thus, we find that the rate at $t_\text{f}=4$ and $N=800$ is in close agreement with the rate at $t_\text{f}=8$ and $N=1600$, indicating convergence. Note that a very good approximation to the reaction rate with less than $5\%$ deviation from the fully converged result can already be obtained with only $50$ beads for propagation times of $2$ or $4$.
\setlength{\tabcolsep}{8pt}
\begin{table}
\caption{Convergence of the rate constant $k$ with respect to propagation time $t_\text{f}$ and number of beads $N$ for the system defined in \eqn{equ:force} with $\nu=4$ and $\epsilon=1$.
}
\label{tab:convergence} 
\begin{tabular}{rccccc}
\toprule
 \multirow{2}{*}{$N$} & \multicolumn{5}{c}{$t_\text{f}$} \\
 \cmidrule{2-6}
      & 1/2     & 1      & 2      & 4      & 8    \\
 \midrule
 50   &  0.1660 & 0.1442 & 0.1155 & 0.1119 & 0.3771 \\   
 100  &  0.1649 & 0.1421 & 0.1178 & 0.1114 & 0.1119 \\
 200  &  0.1644 & 0.1400 & 0.1190 & 0.1135 & 0.1114 \\
 400  &  0.1641 & 0.1404 & 0.1196 & 0.1150 & 0.1135 \\
 800  &  0.1640 & 0.1401 & 0.1199 & 0.1159 & 0.1150 \\
 1600 &  0.1639 & 0.1399 & 0.1201 & 0.1163 & 0.1159 \\
 \bottomrule
\end{tabular}
\end{table}

The DVR results in Fig.~2 have been generated on a grid of lengths 9 and 6 along $x_1$ and $x_2$, discretized by 150 points in each direction \cite{Colbert1992DVR,Shizgal2015}. The first non-zero eigenvalue of the Fokker--Planck operator of a bistable system corresponds to the sum of the forward and backward rate constants \cite{Piserchia2015}. Due to the symmetry of the system defined in \eqn{equ:force}, the forward and backward rate constants are identical, so that $k$ is obtained as half of the smallest non-zero eigenvalue.

The convergence of the instanton rates for the active field theory in \secref{activefield} is somewhat more challenging because the motion along the reaction coordinate is orders of magnitude slower than the motion along some of the other degrees of freedom in the system. Hence, relatively long propagation times of up to $t_\text{f}=60$ for $h=-0.1$ and $\nu=0.25$
are required to reach the long-time limit while a small time step must be used to correctly describe the fast modes. 
In order to converge the instanton rates to within less than $5\%$ error, we therefore used 10000--30000 beads. Due to the efficiency of the method, even calculations with this many beads take less than one day on 32 CPUs. However, it may be possible to reduce the number of required beads by employing different discretization schemes.

\section{Trajectory formalism}
\label{app:trajectory}

The direct calculation of the instanton rate from a discretized path integral [\eqn{equ:k_neqi}] offers high numerical stability but requires diagonalizing a $Nf\times Nf$ matrix, scaling as $\mathcal{O}(N^3)$ with the number of beads. Utilizing the banded structure of $\nabla_\text{s}^2 S_N$ reduces the scaling to $\mathcal{O}(N^2)$. However, for certain large systems requiring many beads for convergence, the rate calculation may still become the bottleneck.
Here, we present an alternative, computationally more efficient approach to calculating the instanton rate within the ``trajectory formalism'', building upon the work of Althorpe and Richardson for Euclidean actions \cite{Althorpe2011ImF,GoldenRPI}.

The key quantity in the definition of the rate constant is the joint probability density [Eqs.~(3),(4)], which we approximate by Laplace's method
\begin{align}
    \label{equ:rhodis}
    \rho(\mat{x}_\text{f},t_\text{f} ; \mat{x}_\text{i},0) \sim [\det( \mathbf{D})]^{-Nf/2}\sqrt{\frac{[\det(\mathbf{J})]^{-1}}{(4\pi \epsilon\, \tau)^f}} \, \eu{-S_N/\epsilon},
\end{align}
where all quantities are evaluated at the instanton, $\bar{\mat{x}}(t)$, connecting $\mat{x}_\text{i}$ to $\mat{x}_\text{f}$ in time $t_\text{f}$. The instanton satisfies $\nabla_\text{r} S_N = 0$, where $\nabla_\text{r} = (\partial/\partial \mat{x}_1,...,\partial/\partial \mat{x}_{N-1})$ 
defines the derivative vector with respect to all intermediate beads. The fluctuations around the instanton are encoded in $\mathbf{J} = 2\tau  \nabla_\text{r}^2 S_N$.   
Alternatively, the probability density in \eqn{equ:rhodis} may be written in analogy to the van-Vleck approximation of the quantum-mechanical propagator \cite{Hunt1981} \footnote{If a system exhibits more than one (dominant) minimum-action pathway, \eqs{equ:rhodis} and \eqref{equ:rhotraj} consist of a sum over all instantons.}
\begin{align}
    \label{equ:rhotraj}
    \rho(\mat{x}_\text{f},t_\text{f} ; \mat{x}_\text{i},0) \sim \sqrt{\frac{C}{(2\pi \epsilon)^f}}
    \, \eu{-S(\mat{x}_\text{i},\mat{x}_\text{f};t_\text{f})/\epsilon},
\end{align}
where the information about the fluctuations around the optimal path is contained in the prefactor
\begin{align}
    \label{equ:C}
    C = \det\left(-\frac{\partial^2 S(\mat{x}_\text{i},\mat{x}_\text{f};t_\text{f})}{\partial \mat{x}_\text{i} \partial \mat{x}_\text{f}}\right) \eu{-\int_{0}^{t_\text{f}} \nabla \cdot \mathbf{\mu} \mat{F}[\bar{\mat{x}}(t)] \rmd t}.
\end{align}
In order for \eqs{equ:rhodis} and \eqref{equ:rhotraj} to be equivalent, the prefactors must be related through
\begin{align}
    \label{equ:CfromJ}
    [\det( \mathbf{D})]^{Nf} (2\tau)^f C = [\det(\mathbf{J})]^{-1} .
\end{align}
We refer to this formalism as the ``trajectory formulation'', because the locations of the $N-1$ intermediate beads are constrained by requiring that they minimize the action (and thus form a classical trajectory) for the given endpoints and time, 
\begin{align}
    \label{equ:implicit}
    \mat{x}_n &=  \bar{\mat{x}}_n \quad \forall n\in [1,N-1] ,
\end{align}
where $\bar{\mat{x}}_n$ are the positions that minimize the discretized action so that $\nabla_\text{r} S_N = 0$.
Hence, the intermediate beads can be integrated out using Laplace's method, so that the action $S(\mat{x}_\text{i},\mat{x}_\text{f};t_\text{f})$ depends explicitly only on the two endpoints, $\mat{x}_\text{i} \equiv \mat{x}_0$ and $\mat{x}_\text{f} \equiv \mat{x}_N$ \cite{Althorpe2011ImF}. 
Therefore, we have $\frac{\partial \mat{x}_\text{i}}{\partial \mat{x}_\text{i}} = \frac{\partial \mat{x}_\text{f}}{\partial \mat{x}_\text{f}} = 1$, $\frac{\partial \mat{x}_\text{i}}{\partial \mat{x}_\text{f}} = \frac{\partial \mat{x}_\text{f}}{\partial \mat{x}_\text{i}} = 0$ and non-trivial derivatives of the intermediate beads with respect to the endpoints.

We now deal with the remaining question of how to evaluate $C$ without calculating $\det(\mathbf{J})$ [\eqn{equ:CfromJ}]. Given the discretized action [\eqn{equ:SN}], a derivative with respect to an intermediate bead is given by
\begin{align}
    \notag \frac{\partial S_N}{\partial \mat{x}_n} &=  
    - \frac{\tau}{2} \left( \frac{1}{\tau} + \nabla \mathbf{\mu}\mat{F}_n \right)^{\!\!\text{T}} \mathbf{D}^{-1}
    \left( \frac{\bar{\mat{x}}_{n+1} - \bar{\mat{x}}_n}{\tau} -  \mathbf{\mu}\mat{F}_n \right) \\
    &+ \frac{1}{2}
    \mathbf{D}^{-1} \left( \frac{\bar{\mat{x}}_{n} - \bar{\mat{x}}_{n-1}}{\tau} - \mathbf{\mu}\mat{F}_{n-1} \right)
    \quad \forall n\in [1,N-1] ,
\end{align}
where $\mat{F}_n \equiv \mat{F}(\bar{\mat{x}}_n)$, and we define $(\nabla\mathbf{\mu}\mat{F})_{jk} = \mu_{jm} \frac{\partial F_m}{\partial x_k}$, where repeated indices are summed over. Taking a second derivative with respect to either of the two endpoints leads to the following system of linear equations
\begin{subequations}
\label{equ:sle}
\begin{align}
    \sum_{n'=1}^{N-1} \sum_{j'=1}^{f} J_{nj,n'j'} \frac{\partial \bar{x}_{n'j'}}{\partial x_{\text{i},k}} &= [\mathbf{D}^{-1} (\mathbb{I}  +\tau \nabla \mathbf{\mu}\mat{F}_0) ] \mat{e}_k \delta_{n1},\\
    \sum_{n'=1}^{N-1} \sum_{j'=1}^{f} J_{nj,n'j'} \frac{\partial \bar{x}_{n'j'}}{\partial x_{\text{f},k}} &=  [\mathbf{D}^{-1} (\mathbb{I}  +\tau \nabla \mathbf{\mu}\mat{F}_{N-1})]^\text{T} \mat{e}_k \delta_{nN-1},
\end{align}
\end{subequations}
where $J_{nj,n'j'}$ are elements of $\mathbf{J}$ with bead indices $n,n' \in [1,N-1]$ and $j,j' \in [1,f]$ indexing the degrees of freedom.
For each degree of freedom $k$ with corresponding unit vector $\mat{e}_k$,  the system of equations is solved for all $\frac{\partial \bar{\mat{x}}_{n}}{\partial x_{\text{i/f},k}}$, vectors containing derivatives of the coordinates of an intermediate bead with respect to the $k\text{th}$ component of endpoint $\mat{x}_\text{i/f}$.

Taking second derivatives of the discretized action with respect to both endpoints under the constraint of \eqn{equ:implicit} results in the relations (see also \Refs{Althorpe2011ImF} and \onlinecite{GoldenRPI})
\begin{subequations}
\label{equ:d2Sdxidxf}
\begin{align}
    \notag \frac{\partial^2 S}{\partial \mat{x}_{\text{i}} \mat{x}_{\text{i}}} &= -\frac{1}{2\tau} \bigg[ \tau [\nabla^2 \mathbf{\mu} \mat{F}_0] [\mathbf{D}^{-1} (\bar{\mat{x}}_1 - \bar{\mat{x}}_0 - \tau \mathbf{\mu} \mat{F}_0)]\\ 
    &+ \left(\mathbb{I} + \tau\nabla\mathbf{\mu}\mat{F}_0\right)^\text{T} \mathbf{D}^{-1} 
    \left(\frac{\partial\bar{\mat{x}}_{1}}{\partial \mat{x}_{\text{i}}}  -\mathbb{I} - \tau\nabla\mathbf{\mu}\mat{F}_0\right) \bigg]  ,\\
    \frac{\partial^2 S}{\partial \mat{x}_{\text{f}} \mat{x}_{\text{f}}} &= \frac{1}{2\tau}  \mathbf{D}^{-1} \left( \mathbb{I} - (\mathbb{I}+\tau \nabla\mathbf{\mu}\mat{F}_{N-1}) \frac{\partial\bar{\mat{x}}_{N-1}}{\partial \mat{x}_{\text{f}}} \right) ,\\
    \frac{\partial^2 S}{\partial \mat{x}_{\text{i}} \mat{x}_{\text{f}}} &= 
    -\frac{1}{2\tau} \frac{\partial\bar{\mat{x}}_{N-1}}{\partial \mat{x}_{\text{i}}}  \left( \mathbb{I} + \tau \nabla\mathbf{\mu}\mat{F}_{N-1} \right)  \mathbf{D}^{-1}
\end{align}
\end{subequations}
where each column of the matrix $\partial\bar{\mat{x}}_n/\partial \mat{x}_{\text{f}}$ is determined by solving \eqs{equ:sle} for a particular $k$, so that $(\frac{\partial\bar{x}_n}{\partial x_{\text{f}}})_{jk} = \frac{\partial\bar{x}_{n,j}}{\partial x_{\text{f},k}}$ and equivalent for derivatives with respect to $\mat{x}_\text{i}$.
The rank 3 tensor in \eqs{equ:d2Sdxidxf} is defined as $(\nabla^2\mathbf{\mu}\mat{F})_{jkl} = \mu_{jm} \frac{\partial^2 F_m}{\partial x_k \partial x_l}$, and $[\nabla^2 \mathbf{\mu} \mat{F}] [\mathbf{D}^{-1} \mat{x}] = (\nabla^2\mathbf{\mu}\mat{F})_{jkl} (\mathbf{D}^{-1} \,\mat{x})_j$.
Together, \eqs{equ:sle} and \eqref{equ:d2Sdxidxf} enable us to compute the prefactor $C$ [\eqn{equ:C}]. %
Notably, the matrices in \eqs{equ:d2Sdxidxf} have dimension $f\times f$, as opposed to $(N-1)f\times (N-1)f$ for $\mathbf{J}$.
We thus transformed the bottleneck of the prefactor calculation from the diagonalization of a large matrix [\eqn{equ:rhodis}] to the solution of a system of linear equations [\eqn{equ:sle}]. The symmetric, banded, positive-definite properties of $\mathbf{J}$ permit the use of efficient solution algorithms, reducing the scaling of the computational costs to $\mathcal{O}(N)$. %

The trajectory formalism developed above can be used to approximate probability densities and related quantities of interest. However, as laid out in \appref{timetranslation}, the fluctuation matrix obtained in the calculation of rate constants is ill-defined because of the time-translation symmetry of the instanton pathway. In the discretized path-integral formalism [\eqn{equ:rhodis}], this symmetry manifests itself as a zero eigenvalue of $\mathbf{J}$, which can be readily removed.  The trajectory formalism lacks a similar solution. 
However, we can break time-translation symmetry by splitting the instanton into two shorter segments at an arbitrary point $\mat{x}_\text{m}$ along the trajectory \cite{InstReview,Getfert2010}, using the composition property $\rho(\mat{x}_\text{f},t_\text{f} ; \mat{x}_\text{i},0) = \int \rho(\mat{x}_\text{f},t_\text{f} ; \mat{x}_\text{m},t_\text{m}) \rho(\mat{x}_\text{m},t_\text{m} ; \mat{x}_\text{i},0) \, \rmd \mat{x}_\text{m}$.
The integral to reconnect the probabilities can again be taken by Laplace's method, so that the prefactor of the total probability density is given by
\begin{multline}
    \label{equ:comp}
    C = C_\text{a} C_\text{b} \\ \times \left[ \det\left(\frac{\partial^2 S(\mat{x}_\text{i},\mat{x}_\text{m};t_\text{m})}{\partial \mat{x}_\text{m}^2}
    + \frac{\partial^2 S(\mat{x}_\text{m},\mat{x}_\text{f};t_\text{f} - t_\text{m})}{\partial \mat{x}_\text{m}^2}
    \right) \right]^{-1} ,
\end{multline}
where the prefactors of the shorter segments are
\begin{align}
    C_\text{a} &= \det\left(-\frac{\partial^2 S(\mat{x}_\text{i},\mat{x}_\text{m};t_\text{m})}{\partial \mat{x}_\text{i} \partial \mat{x}_\text{m}}\right) \eu{-\int_{0}^{t_\text{m}} \nabla \cdot \mathbf{\mu} \mat{F}[\mat{x}(t)] \rmd t},\\
    C_\text{b} &= \det\left(-\frac{\partial^2 S(\mat{x}_\text{m},\mat{x}_\text{f};t_\text{f}-t_\text{m})}{\partial \mat{x}_\text{m} \partial \mat{x}_\text{f}}\right) \eu{-\int_{t_\text{m}}^{t_\text{f}} \nabla \cdot \mathbf{\mu} \mat{F}[\mat{x}(t)] \rmd t} ,
\end{align}
and the connection piece can also be evaluated with \eqs{equ:d2Sdxidxf} and \eqref{equ:sle}.

However, fully reconnecting the segments would reintroduce the problematic time-translation mode. Instead, we exploit the fact that this mode aligns with the instanton velocity along the path. Therefore, we can \rred{use the Faddeev--Popov method \cite{Faddeev1967}} and separate the coordinates of the splitting point $\mat{x}_\text{m}$ into the mode parallel to the instanton $q_\text{m}$ and the modes orthogonal to it $\mat{Q}_\text{m}$. Integration is then performed only over the latter. Applying this approach to \eqn{equ:def_rate} yields the following equation for the instanton rate constant
\begin{multline}
    \label{equ:pretraj}
    k \sim \lim_{\mathcal{T}\rightarrow\infty} \frac{\sqrt{C_\text{a} C_\text{b}}}{(2\pi\epsilon)^f} \, \frac{\dot{\bar{q}}_\text{m}}{\bar{p}_\text{f}} \, \sqrt{\frac{(2\pi\epsilon)^{f-1}}{\Omega_\text{m}}} \, \sqrt{\frac{(2\pi\epsilon)^{f-1}}{\Omega^\ddagger}}\\ 
    \times \int_0^\mathcal{T} 
    \bar{\mat{p}}^\text{T}_\text{f} \mathbf{D} \, \bar{\mat{p}}_\text{f} \, \eu{-S(\mat{x}_\text{R}, \mat{x}^\ddagger, t_\text{f})/\epsilon} \, \rmd t_\text{m},
\end{multline}
where $\Omega_\text{m} = \det\left(\frac{\partial^2 S(\mat{x}_\text{R},\mat{x}_\text{m};t_\text{m})}{\partial \mat{Q}_\text{m}^2} + \frac{\partial^2 S(\mat{x}_\text{m},\mat{x}^\ddagger;t_\text{f} - t_\text{m})}{\partial \mat{Q}_\text{m}^2}\right)$ originates from the composition integral over the orthogonal modes.
The Laplace integration over the dividing surface leads to the term $\Omega^\ddagger = \det\left(\frac{\partial^2 S(\mat{x}_\text{R},\mat{x}^\ddagger;t_\text{f})}{\partial \mat{s}^2}\right)$, where the reaction coordinate along $\bar{\mat{p}}_\text{f}$ has been projected out and the remaining $f-1$ coordinates parameterizing the dividing surface are combined in $\mat{s}$.   

The velocity along the instanton at the splitting point, $\dot{\bar{q}}_\text{m}$, appears because we transformed the integral over $q_\text{m}$ to a collective time integral.
In analogy to the prefactor in the discretized path-integral formulation, we find that $C_\text{b} \propto \bar{\mat{p}}_\text{f}^2$.
Consequently, the term $C_\text{b}/\bar{\mat{p}}_\text{f}^2$ can be factored out of the integral and evaluated at the instanton because it varies slowly compared to the exponential even though  $\bar{\mat{p}}_\text{f}\rightarrow 0$ in the infinite-time limit \cite{BenderBook}. 
By noting that the prefactor in the integrand of \eqn{equ:pretraj} is the instanton energy [\eqn{equ:momentum}] and recognizing that the action of a path connecting distinct points in zero time is infinite, the remaining collective time integral can be solved by substitution. We thus arrive at the nonequilibrium instanton formula for the rate constant within the trajectory formalism
\begin{align}
    \label{equ:ktraj}
    k \sim %
    \frac{1}{2\pi} \, \frac{\dot{\bar{q}}_\text{m}}{\bar{p}_\text{f}} \, \sqrt{\frac{C_\text{a} C_\text{b}}{\Omega_\text{m} \Omega^\ddagger}} \,
    \eu{-S/\epsilon} ,
\end{align}
whose numerical evaluation scales as $\mathcal{O}(N)$.

In practice, achieving a desired numerical accuracy for $C$, calculated from derivatives with respect to opposite ends of the path, often requires more beads than the convergence of the connection pieces, obtained from derivatives with respect to the same end [\eqn{equ:d2Sdxidxf}].
Therefore, computing $C$ from $\det(\mathbf{J})$, as shown in \eqn{equ:CfromJ}, may proof advantageous.
The efficiency of the method can be further enhanced by splitting up the paths into smaller segments, thus iteratively applying \eqn{equ:comp} in analogy to \Refx{DaC}. %
In this way, the diagonal parts of \eqs{equ:d2Sdxidxf} can be used to reconnect the paths while the prefactors $C_\text{a},C_\text{b},...$ are obtained by diagonalizing the smaller $\mathbf{J}$ matrices of the shorter segments. 
The independent calculations for individual segments are trivially parallelizable, leading to an approach that is both efficient and numerically robust. 
Alternatively, the components of the prefactor in \eqn{equ:ktraj} may be computed by numerically solving matrix Riccati or Gelfand--Yaglom equations, as suggested for large-deviation functions \cite{Schorlepp2021,Grafke2023,Schorlepp2023,Bouchet2022,Lehmann2003}. 
This highlights the broader applicability of our methods beyond reaction rate theory.

The extension of the instanton rate in the trajectory formalism [\eqn{equ:ktraj}] to field-theoretical descriptions, analogous to \eqn{equ:k_neqi_fieldT}, is straightforward. 
The only differences to \eqn{equ:ktraj} are that: (i) the $f$ zero modes arising from the translation symmetry at the TS must be excluded from the integral over the dividing surface and thus projected out of $\Omega^\ddagger$ using the corresponding normal modes at the TS\@, and (ii) the volume term $\left(\frac{B_x}{2\pi\epsilon\kappa}\right)^{\!f/2}$ from the integration over these translation modes must be included.

\section{Connection to Kramers--Langer theory at equilibrium}
\label{app:kramers}

The transition-state theory estimate of the rate constant at equilibrium is given by Kramers--Langer theory (KLT) \cite{Haenggi1990,Kramers1940,Langer1969,Langer1973}
\begin{figure}
    \centering
    \includegraphics[width=0.48\textwidth]{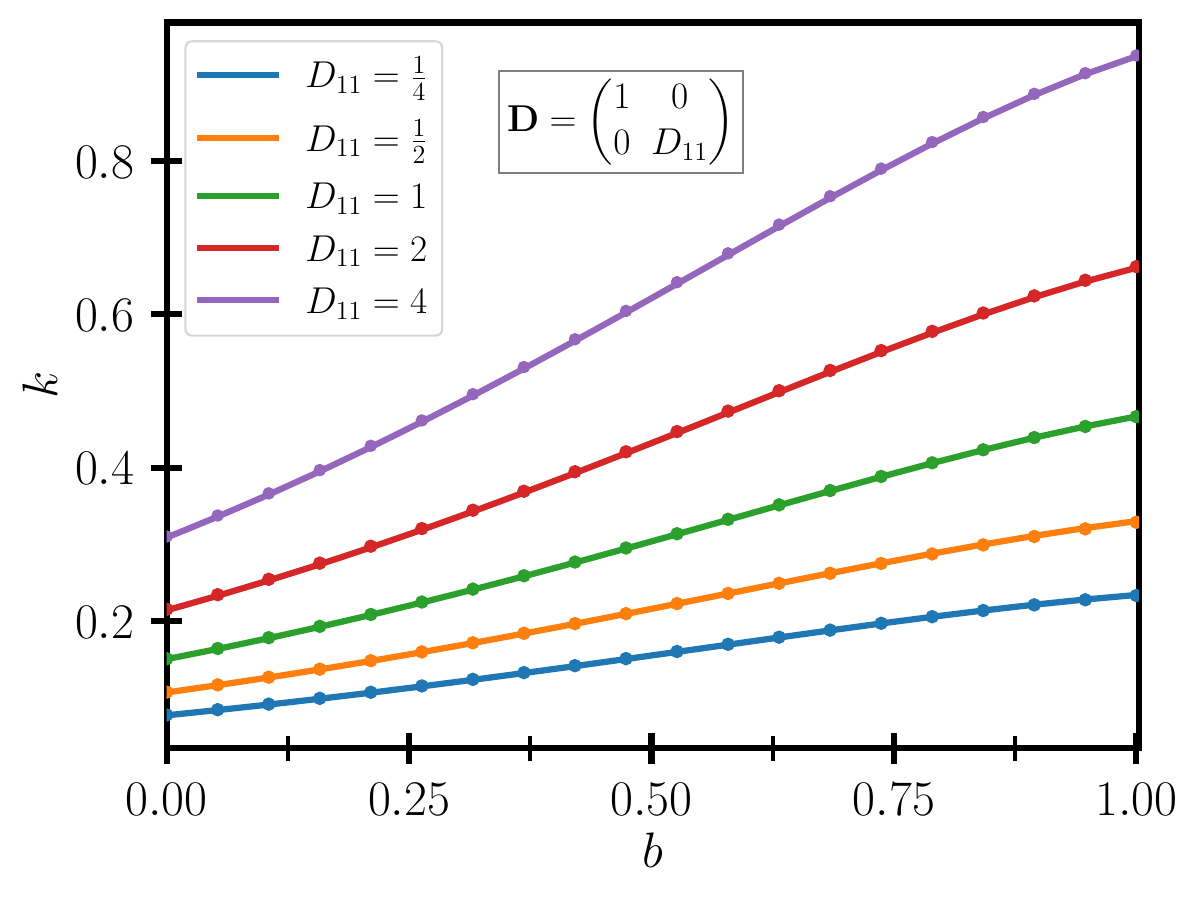}
    \caption{KLT (solid lines) and instanton (dots) rate constants for the potential defined in \eqn{equ:potentialSI} over a range of biases, $b$, and diffusion anisotropies with $\beta = \epsilon = 1$. 
    }
    \label{fig:KLTrates}
\end{figure}
\begin{figure}
    \centering
    \includegraphics[width=0.48\textwidth]{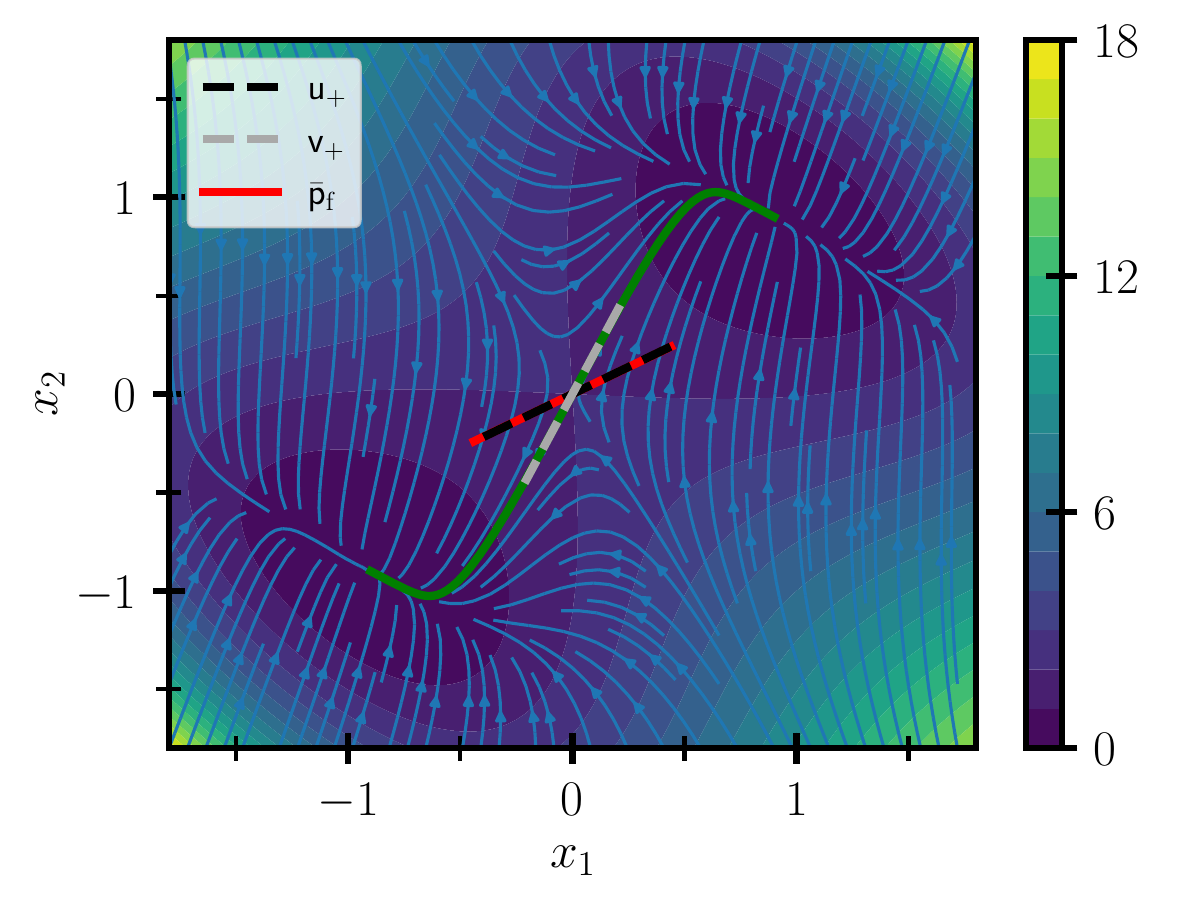}
    \caption{Color map of the potential defined below \eqn{equ:force}. The diffusion is anisotropic because $\mathbf{D} = \frac{1}{2} \begin{pmatrix} 2 & 1\\ 1 & 8 \end{pmatrix} \neq \mathbb{I}$. The stream plot depicts $\mathbf{\mu} \mat{F}$, where $\mat{F}$ is conservative ($\nu=0$). The instanton is shown by the green line along with the direction of the canonical momentum $\bar{\mat{p}}_\text{f}$ at the TS and the eigenvectors $\mat{u}_{+},\mat{v}_{+}$ associated with the only positive eigenvalues of $\nabla \mat{F}(\mat{x}^\ddagger) \mathbf{\mu}$ and $\nabla\mathbf{\mu}\mat{F}(\mat{x}^\ddagger)$, respectively. 
    }
    \label{fig:anisotropy}
\end{figure}
\begin{align}
    \label{equ:Kramers}
    k_\text{KLT} &= \frac{\lvert\lambda_\text{b}\rvert}{2\pi} \, \sqrt{\frac{\det(\nabla^2 V_\text{R})}{\lvert\det(\nabla^2 V^\ddagger)\rvert}}\, \eu{-\beta V^\ddagger} ,
\end{align}
where $V^\ddagger = V(\mat{x}^\ddagger) - V(\mat{x}_\text{R})$ is the barrier height, $\nabla^2 V_\text{R}$ and $\nabla^2 V^\ddagger$ are the Hessians at the reactant and TS configurations, and $\lambda_\text{b}$ is the sole negative eigenvalue of $\mathbf{\mu }\nabla^2 V^\ddagger$ associated with the motion across the barrier. 
In \Refx{Getfert2010}, it has been explicitly shown that instanton theory is equivalent to KLT in one-dimensional systems at equilibrium. Given the equivalence of the Fokker--Planck representation of the rate, from which KLT is commonly derived, and the path-integral representation underlying instanton theory, it is clear that our instanton theory correctly reduces to multi-dimensional KLT in the limit of thermodynamic equilibrium (purely conservative forces). We illustrate this point in Fig.~\ref{fig:KLTrates}, where we compare KLT and instanton results for the potential 
\begin{align}
    \label{equ:potentialSI}
    V(\mat{x}) &= \frac{A}{B^4}\left(x_{+}^2-B^2\right)^{\!2} + x_{-}^2 - b x_{+}, 
\end{align}
using the parameters defined below \eqn{equ:force}. 
The rate constants are shown for a range of different biases $b$ and diffusion anisotropies. 
As expected, the two rate estimates agree to within numerical precision.

Beyond the value of the rate constant, Fig.~\ref{fig:anisotropy} further elucidates the connection between the KLT and instanton pictures of the reaction mechanism. It can be seen that the instanton velocity (pointing along the path) aligns with the eigenvector $v_{+}$ associated with the sole positive eigenvalue of $\nabla \mathbf{\mu} \mat{F}(\mat{x}^\ddagger)$ at the TS\@, which indicates the direction of the reactive flux. 
Instead, the canonical momentum $\bar{\mat{p}}_\text{f}$ aligns with the eigenvector $u_{+}$ associated with the only positive eigenvalue of $\nabla \mat{F}(\mat{x}^\ddagger) \mathbf{\mu}$. This eigenvector points along the unstable diffusive normal mode at the TS\@, which is orthogonal to the stochastic separatrix and thus defines the reaction coordinate \cite{Berezhkovskii2004}. 
Instanton theory thus recovers these concepts from KLT at equilibrium and extends them to far-from-equilibrium regimes.

\section{Active field theory}
\label{app:activefield}

\begin{figure}
    \centering
    \includegraphics[width=0.48\textwidth]{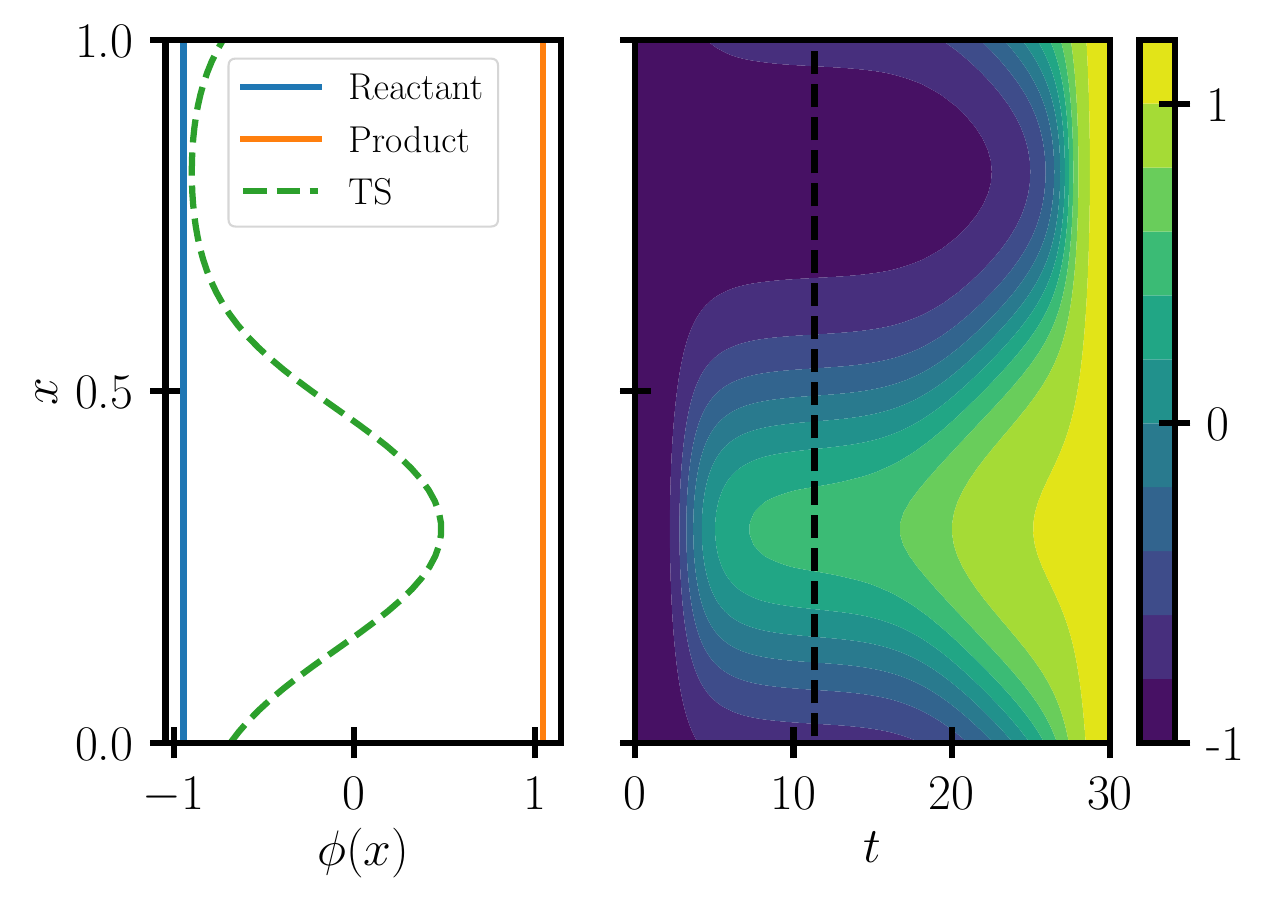}
    \caption{(Left) Field configuration at the stationary points of the potential [\eqn{equ:fieldpotential}] associated with the stochastic field theory defined in \secref{activefield} ($\nu=0$) with $h=0.1$.
    (Right) Instanton of the same system, where the color map indicates the field configuration along the path as a function of space and time.
    The black dashed line highlights the TS configuration or critical nucleus. 
    }
    \label{fig:interface}
\end{figure}

At equilibrium, the instanton rate equation [\eqn{equ:k_neqi_fieldT}] for active field theories reduces to the corresponding KLT rate \cite{Langer1969,Ekstedt2022a,Cottingham1993,Simeone2023}
\begin{align}
    \frac{k_\text{KLT}}{\mathcal{V}} &= \left(\frac{B_x}{2\pi\epsilon\kappa}\right)^{\!\! f/2} \frac{\lvert\lambda_\text{b}\rvert}{2\pi} \, \frac{Z^\ddagger}{Z_\text{R}}\, \eu{-\beta V^\ddagger} .
\end{align}
For a $\phi^4$ theory with a potential functional of the form 
\begin{align}
    \label{equ:fieldpotential}
    V[\phi(x)] &= \int \left[ \frac{\kappa}{2} \, \lvert\nabla \phi \rvert^2 + W(\phi) \right] \rmd \mat{x},
\end{align}
the fluctuation factors at the reactant configuration, $\phi(\mat{x})=\phi_\text{R}(\mat{x})$, and the TS configuration, $\phi(\mat{x})=\phi^\ddagger(\mat{x})$, are given by
\begin{subequations}
\label{equ:pfs}
\begin{align}
    Z^\ddagger &= \left\lvert\text{det}'\left(-\kappa\Delta + \frac{\partial^2 W}{\partial \phi^2} \right)_{\!\!\phi(\mat{x})=\phi^\ddagger(\mat{x})}\right\rvert^{-1/2} ,\\
    Z_\text{R}         &= \left[\det\left(-\kappa\Delta  + \frac{\partial^2 W}{\partial \phi^2}  \right)_{\!\!\phi(\mat{x})=\phi_\text{R}(\mat{x})}\right]^{\!-1/2} ,
\end{align}
\end{subequations}
where $\lambda_\text{b}$ is the single negative eigenvalue of $\mu [-\kappa\Delta + \partial^2 W/\partial \phi^2]$ at the TS\@.
The translation symmetry of the interface formed at the TS leads to zero eigenvalues in the Hessian.
The prime in the definition of $Z^\ddagger$ indicates that these zero eigenvalues have been removed from the determinant. 

From the stationarity condition at the TS\@,  $\delta V/\delta \phi = 0$, it follows that %
$E_x = \kappa \lvert\nabla \phi\rvert^2/2 - W(\phi)$ is constant (independent of $\mat{x}$) along $\phi^\ddagger(\mat{x})$. 
We can therefore add a zero under the integral in \eqn{equ:fieldpotential}, leading to
\begin{align}
    \label{equ:addzero}
    \notag V[\phi^\ddagger(x)] &= \int \left[ \frac{\kappa}{2} \, \lvert\nabla \phi^\ddagger \rvert^2 + W(\phi^\ddagger) + E_x(\phi^\ddagger) - E_x \right] \rmd \mat{x} .\\
               &= B_x + V[\phi_{>}(\mat{x})] , %
\end{align}
Here, we used that,
in the absence of finite-size effects, $\phi^\ddagger(\mat{x})$ is dominated by the less stable (uniform) state $\phi_{>}$ with only a small nucleus (or bubble) extending towards the lower-energy state (see Fig.~\ref{fig:interface}), so that $V[\phi_{>}(\mat{x})] = -\int E_x \, \rmd \mat{x}$.  
In the common case of an exothermic nucleation process, one obtains the simple relationship $B_x =  V^\ddagger = V[\phi^\ddagger(\mat{x})] - V[\phi_\text{R}(\mat{x})]$,
in agreement with previous work on nucleation rates at equilibrium \cite{Cottingham1993,Simeone2023}.
This, however, does not hold out of equilibrium, where the TS is not a stationary point of a potential. 

In practice, we represent the field on a grid. Using \eqn{equ:fieldpotential} in one dimension as an example, the discretized potential on a grid of $N_x$ points spaced by $\xi$ is given by
\begin{multline}
    \label{equ:Vdis}
    V(\varphi_0,...,\varphi_{N_x-1}) =\\ \xi \sum_{i=0}^{N_x-1} \frac{\kappa}{2} \left( \frac{\varphi_{i+1} - \varphi_{i}}{\xi} \right)^{\!\!2} + W(\varphi_i) ,
\end{multline}
where $\varphi_i \equiv \phi(x_i)$,  
and $\varphi_0 = \varphi_N$ because we employ periodic boundary conditions. 
The optimization of stationary points and evaluation of the fluctuation terms [\eqs{equ:pfs}] can then be carried out by taking derivatives with respect to the $\varphi_i$ and correctly accounting for $\xi$. This procedure is analogous to the discretization of the instanton in time [\eqn{equ:SN}].   
The corresponding discretized instanton action in field space [\eqn{equ:fieldaction}] is given by 
\begin{align}
    \label{equ:SNfield}
    S_N(\phi_0,..., \phi_N;t_\text{f}) = \frac{\tau\xi}{4} \sum_{n=0}^{N-1} \left(\frac{\phi_{n+1} - \phi_n}{\tau} - \mu F(\phi_{n})\right)^{\!\! 2}  ,
\end{align}
where each bead $\phi_n = ( \varphi_0,...,\varphi_{N_x-1} )$ corresponds to a field configuration, and $F(\phi_{n}) = - \delta V(\phi_{n})/\delta \phi + F_\text{nc}(\phi_{n})$ is the sum of all conservative and non-conservative forces.
The instanton connects $\phi_0 \equiv \phi_\text{R}$ to $\phi_N \equiv \phi^\ddagger$. It is obtained by minimizing \eqn{equ:SNfield} with respect to all intermediate beads $(\phi_1,...,\phi_{N-1})$.

The path-integral representation of the rate is given by
\begin{multline}
    \label{equ:PI_field}
    k \, Z_\text{R} =
    \lim_{N,N_x\rightarrow \infty}  \left(\frac{\xi}{4\pi\epsilon \tau}\right)^{\!\!N N_x/2}\\ \times  \int \dot{\phi}_\sigma(t_\text{f}) \, \eu{-S_N(\phi_0,..., \phi_{N};t_\text{f})/\epsilon} \, \rmd \phi_1 ... \rmd \phi_{N-1} \rmd \mat{s} ,   
\end{multline}
where $\dot{\phi}_\sigma(t_\text{f})$ is the velocity orthogonal to the dividing surface and the generalization to $f$ dimensions is straightforward.
Carrying out a Laplace approximation in analogy to \secref{derivation} then leads to the instanton rate for active field theories in \eqn{equ:k_neqi_fieldT}. 
The fluctuations around the instantons are contained within the determinant of $\nabla^2_\phi S_N$, where $\nabla_\phi = ( \partial/\partial \phi_1,...,\partial/\partial \phi_{N-1}, \partial/\partial \mat{s} )$ is the derivative vector with respect to all intermediate beads and the parameters of the field-space dividing surface $\mat{s}$.
However, in addition to the time-translation mode, also the modes corresponding to translation of the interface must be projected out via the zero eigenvectors of $\partial F(\phi^\ddagger)/\partial\phi$. 
As discussed in \secref{derivation}, the reaction coordinate at the TS can be projected out via $\bar{p}_\text{f} = \frac{1}{\xi}\frac{\partial S_N}{\partial \phi_N}$.

\begin{figure}
    \centering
    \includegraphics[width=0.48\textwidth]{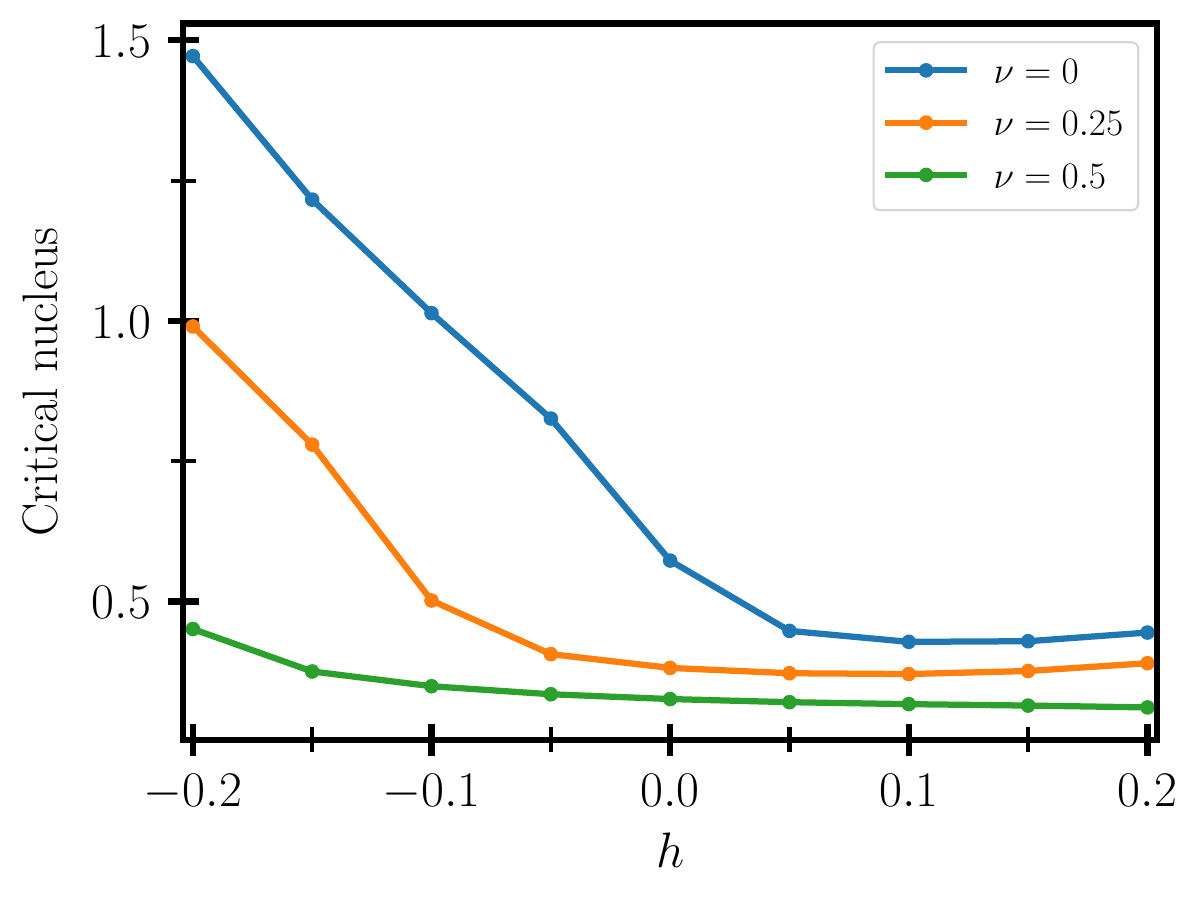}
    \caption{Size of the critical nuclei in the phase transition of the active field theory in \secref{activefield} over a range of applied fields $h$ and nonequilibrium driving strengths $\nu$.
    The size of the critical nucleus is measured by  ${\int [\phi^\ddagger(x) - \phi_\text{R}(x)] \, \rmd x / [\text{max}(\phi^\ddagger) - \text{min}(\phi^\ddagger)]}$.
    }
    \label{fig:nuclei}
\end{figure}
In Fig.~\ref{fig:interface}, we illustrate the KLT and instanton pictures of a transition in field space. As before, the instanton smoothly connects the stationary points through a trajectory in time. It can be seen that, at the TS\@, part of the field extends from the reactant to the product configuration, forming a ``critical nucleus'' or an interface between the states. The nucleus is smaller than the box in which the field is defined. Given the periodic boundary conditions, the nucleus can form anywhere in the box without a change in the activation barrier, giving rise to the previously mentioned continuous translation symmetry at the TS\@. 
\begin{figure*}[t!]
    \centering
    \includegraphics[width=1.0\textwidth]{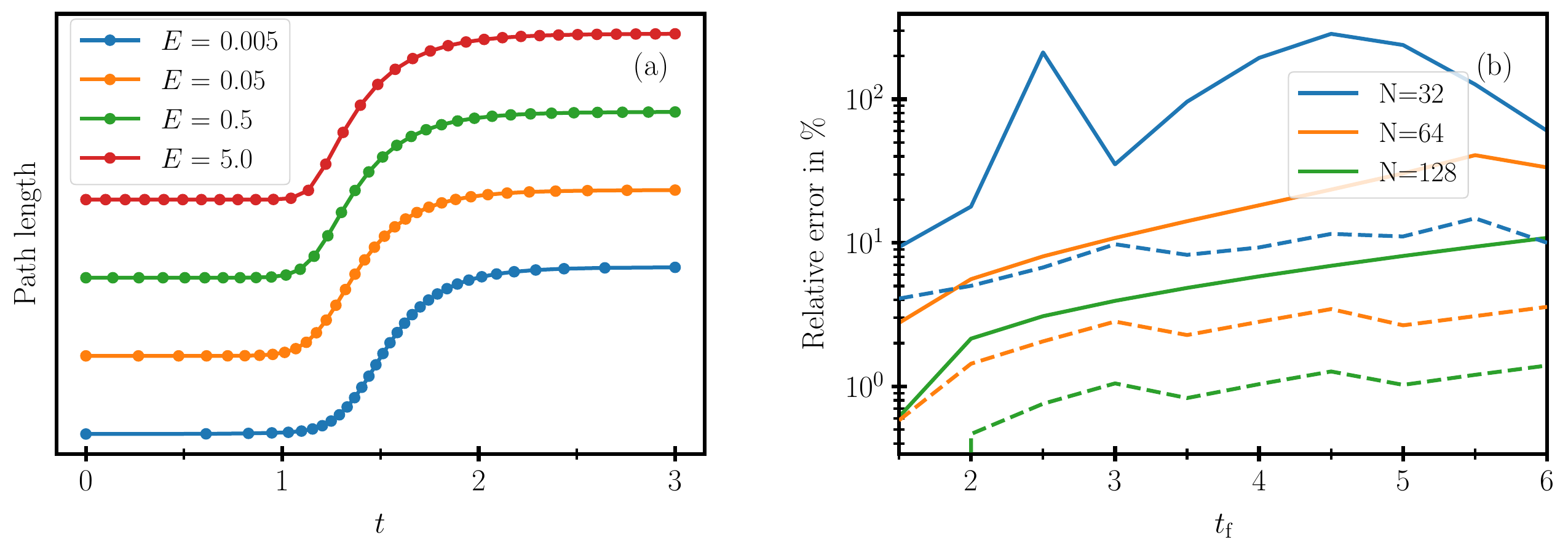}
    \caption{\red{
    Instanton calculations for the system defined in \eqn{equ:force} with $\nu=0$.  
    (a) Path length as function of time $t$ for optimized instantons with $N=32$ beads, a total propagation time of $t_\text{f}=3$ and variable time step with different choices for the parameter $E$. The smaller the value of $E$, the fewer beads are located near the fixed points of the force. For $E=5$, the time intervals are virtually identical. 
    (b) Error of the instanton rate constants relative to the KLT rate constant as a function of total propagation time $t_\text{f}$ for different number of beads. Solid lines correspond to calculations with equal time intervals whereas dashed lines are obtained from calculations with variable time intervals, using values for $E$ in the range $5\cdot 10^{-1} - 5\cdot 10^{-4}$. 
    } 
    }
    \label{fig:varstep}
\end{figure*}

Fig.~\ref{fig:nuclei} explores how the size of the critical nuclei, which need to be formed to facilitate a transition, varies with the strength of the applied field $h$ and the nonequilibrium driving $\nu$. 
It can be seen that with increasing driving strength, the required size of the critical nucleus decreases.  
This directly correlates with the rate enhancement due to nonequilibrium driving observed in \fig{nucleation}.

It is worth pointing out that, for the active field theory defined in \secref{activefield}, the backward transitions proceed via a uniform TS when $\nu\neq 0$ because of the finite size of the box \cite{Zakine2023}. For our purposes, this is not a problem since a meaningful action can still be extracted.

\red{
\section{Rate calculation with variable time steps}
\label{app:timestep}

Beads tend to accumulate around the fixed points of the force, which is due to the kink shape of the instanton in Fig.~\ref{fig:doublekink}. Several methods have been devised to distribute the beads in a more uniform way, either by transforming from physical time to an arc length coordinate \cite{Heymann2008,Heymann2008a,Zakine2023} or by lifting the requirement of equal time steps \cite{Zhou2008,Rommel2011grids,Cvitas2016instanton}, typically choosing longer time steps close to the fixed points and shorter time steps where the system configuration changes fastest.

Especially the latter strategy is easily incorporated into the presented formalism by following \Refx{Rommel2011grids}. The instanton then corresponds to the minimum of the discretized action
\begin{multline}
    \label{equ:SNtau}
    \tilde{S}_N = \frac{1}{4} \sum_{n=0}^{N-1} \left(\frac{\mat{x}_{n+1} - \mat{x}_n}{\sqrt{\tau_{n+1}}} - \sqrt{\tau_{n+1}} \,\mathbf{\mu} F(\mat{x}_n)\right)^{\!\!\text{T}}\\  
    \times \mathbf{D}^{-1} \left(\frac{\mat{x}_{n+1} - \mat{x}_n}{\sqrt{\tau_{n+1}}} - \sqrt{\tau_{n+1}} \, \mathbf{\mu} F(\mat{x}_n)\right) ,
\end{multline}
with variable time intervals $\tau_n>0$. 
The rate expression from \eqn{equ:k_neqi} generalized for flexible time intervals is given by 
\begin{align}
    \label{equ:k_tau}
    k &\sim \frac{1}{2\pi\tilde{p}_\text{f}} \, \sqrt{\tilde{B}_t} \, \frac{\prod_{n=1}^{Nf} \lambda^\text{R}_n}{\prod_{n=3}^{Nf} \lambda^\text{I}_n}
    \, \eu{-\tilde{S}_N/\epsilon},
\end{align}
where $(\lambda_n^\text{R/I})^2$ are the eigenvalues of the time-weighted second-derivative matrix of the action with blocks $\mathbf{H}_{mn} = \frac{\partial^2 \tilde{S}_N}{\partial \mat{x}_m \partial \mat{x}_n}  / \sqrt{\tau_n \tau_m}$ evaluated at the reactant and instanton configuration respectively, where $n,m\in [1,N]$. The reactant path is collapsed at $\mat{x}_\text{R}$, which can be used to simplify the calculation of the eigenvalues $(\lambda_n^\text{R})^2$ \cite{Rommel2011grids}. As in \eqn{equ:k_neqi}, the time-translation mode and the reaction coordinate are excluded from the calculation of the fluctuation factor at the instanton.
Further, we defined the time-weighted canonical momentum at the endpoint $\tilde{\mat{p}}_\text{f} = \frac{\partial \tilde{S}_N}{\partial \mat{x}_N} / \sqrt{\tau_N}$ with norm $\tilde{p}_\text{f} = \lVert \tilde{\mat{p}}_\text{f} \rVert$, and we indicate that $\tilde{B}_t$ must be evaluated on the chosen (non-uniform) time grid.

In analogy to \Refx{Rommel2011grids}, a possible choice for the $\tau_n$ can be derived from the conservation of the energy along the instanton. Consequently, $\lVert \dot{\mat{x}}(t) \rVert = \sqrt{4E + [\mathbf{\mu} \mat{F}(\mat{x}(t))]^2}$ and upon discretization $\tau_{n+1} = \frac{\Delta x}{\sqrt{4E + [\mathbf{\mu} \mat{F}(\mat{x}_n)]^2}}$, where $E$ can be chosen so as to achieve the desired distribution of beads and the constant $\Delta x$ is fixed by the relation $\sum_{n=1}^{N} \tau_n= t_\text{f}$. 
In the process of converging the rate constant, instantons with different $N$ are optimized, as shown in Table~\ref{tab:convergence}. Based on a previously optimized path (e.g., with lower $N$), a good estimate for $E$ can be obtained as the average of the instantaneous energy $E_n = \bar{\mat{p}}_n^\text{T} \mathbf{D} \bar{\mat{p}}_n + \bar{\mat{p}}_n^\text{T} \mathbf{\mu} \mat{F}(\bar{\mat{x}}_n)$ evaluated at each bead along the path. %
In Fig.~\ref{fig:varstep}(a), we show optimized instantons at different values of $E$ initialized from a straight line with equidistant segments connecting reactant minimum and TS\@. It can be seen that smaller values of $E$ lead to a larger number of beads in the transition region because the time steps in the center of the path shorten compared to the time steps close to the endpoints. Note that it is crucial to project out the time-translation mode in the optimization of these instantons with flexible time step. 

From Fig.~\ref{fig:varstep}(b), it can be seen that a decent choice for $E$ can significantly enhance the rate of convergence of the instanton results with respect to the number of beads.    
However, it is worth noting that \eqn{equ:k_tau} converges to the same result in the $N\rightarrow \infty$ limit irrespective of the choice for the $\tau_n$. %
In the spirit of the geometric minimum action method \cite{Heymann2008,Heymann2008a,Zakine2023}, also segments of equal arc length can be obtained by solving the system of equations $\frac{\partial \tilde{S}_N}{\partial \mat{x}_n}=0$ for the $\tau_n$ under the constraint $\sum_{n=1}^{N} \tau_n= t_\text{f}$ based on an initial path with equidistant segments. However, this typically does not lead to the fastest convergence of the rate constant because very few beads are placed close to the fixed points \cite{Rommel2011grids}. 

\section{State-dependent diffusion}
\label{app:multiplicative}

Here, we show that the proposed instanton method extends straightforwardly to the case of state-dependent diffusion, where the dynamics are governed by an overdamped Langevin equation with multiplicative (position-dependent) noise
\begin{align}
    \label{equ:langevin_mult}
    \dot{\mat{x}}(t) = \mathbf{\mu}(\mat{x}) \mat{F}[\mat{x}(t)] +  \sqrt{2 \epsilon} \, \mathbf{\Lambda}(\mat{x}) \,\eta(t).
\end{align}
Both the diffusion matrix, $\mathbf{D}(\mat{x}) = \mathbf{\Lambda}(\mat{x})\mathbf{\Lambda}^{\!\text{T}}(\mat{x})$, and mobility matrix, $\mathbf{\mu}(\mat{x}) = \beta \epsilon \mathbf{D}(\mat{x})$, depend on position, and we again assume $\mathbf{D}(\mat{x})$ to be non-singular. The derivatives of the discretized action [\eqn{equ:SN}] thus also involve derivatives of the diffusion and mobility matrices. However, the general procedure of deriving \eqn{equ:k_neqi} as well as its numerical evaluation remains the same. Accounting for the position-dependence of the diffusion matrix in the path measure then leads to the following definition of the fluctuation determinant  $\Sigma = \det'[2\tau \nabla^2_\text{s} S_N] \prod_{n=1}^{N}\text{det}[\mathbf{D}(\mat{x}_n)]$. \rred{Here, we employ the It\^o convention for the interpretation of the stochastic differential equation as well as for the corresponding path integral. The instanton method, however, can also be used with other discretization conventions \cite{Pirey2022}.}
\begin{figure*}[t!]
    \centering
    \includegraphics[width=\textwidth]{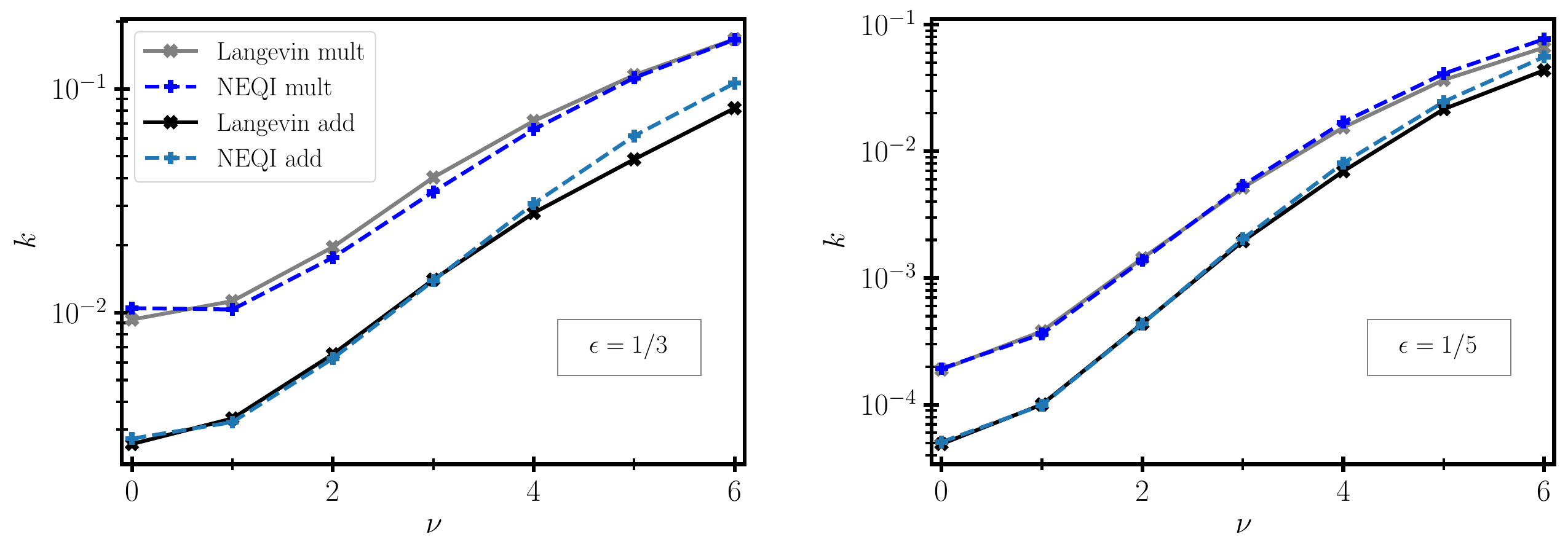}
    \caption{\red{
    Rate constants for the model with multiplicative noise defined in the text for two different noise strengths $\epsilon$ over a range of driving strengths $\nu$ computed from nonequilibrium instanton theory (NEQI) and a numerically exact benchmark. Lines are drawn as guide for the eye. The benchmark results were obtained through numerical integration of the Langevin dynamics in \eqn{equ:langevin_mult} in It\^o convention using the Euler--Maruyama method with a time step of $\Delta t = 5\cdot 10^{-3}$.
    The rate constant is computed as the inverse of the mean first passage time of 500 trajectories initialized at random positions within the reactant region. Each data point constitutes the mean of five of these calculations initialized with different random seeds. The standard deviation is smaller than the symbols indicating the data points.    
    For comparison, the instanton and exact rate constants are given for the same system with additive noise, where $\mathbf{\mu} = \mathbb{I}$. 
    } 
    }
    \label{fig:mult_rates}
\end{figure*}

In order to illustrate the performance of our instanton method for systems with multiplicative noise, we consider the two-dimensional model system defined in \eqn{equ:force} with the position-dependent mobility ${\mathbf{\mu}(\mat{x}) = \begin{pmatrix} 1+(x_0+x_1)^2 & 0\\ 0 & 1+x_1^2 \end{pmatrix}}$. In Fig.~\ref{fig:mult_rates}, we compare the instanton rate constants with numerically exact results from direct integration of \eqn{equ:langevin_mult} for two different values of the noise strength over a range of nonequilibrium driving strengths. As in the case of additive noise, the instanton results are in excellent agreement with the benchmark for sufficiently weak noise. Again, we observe that the error of the instanton approximation increases at very strong driving, where for $\nu=6$ the instanton action $S\approx 0.384$ becomes comparable to the noise strength and the assumptions of weak noise and time-scale separation are less justified.   

}

\rred{
\section{Underdamped dynamics}
\label{app:ud}

\begin{figure}[b]
    \centering
    \includegraphics[width=0.48\textwidth]{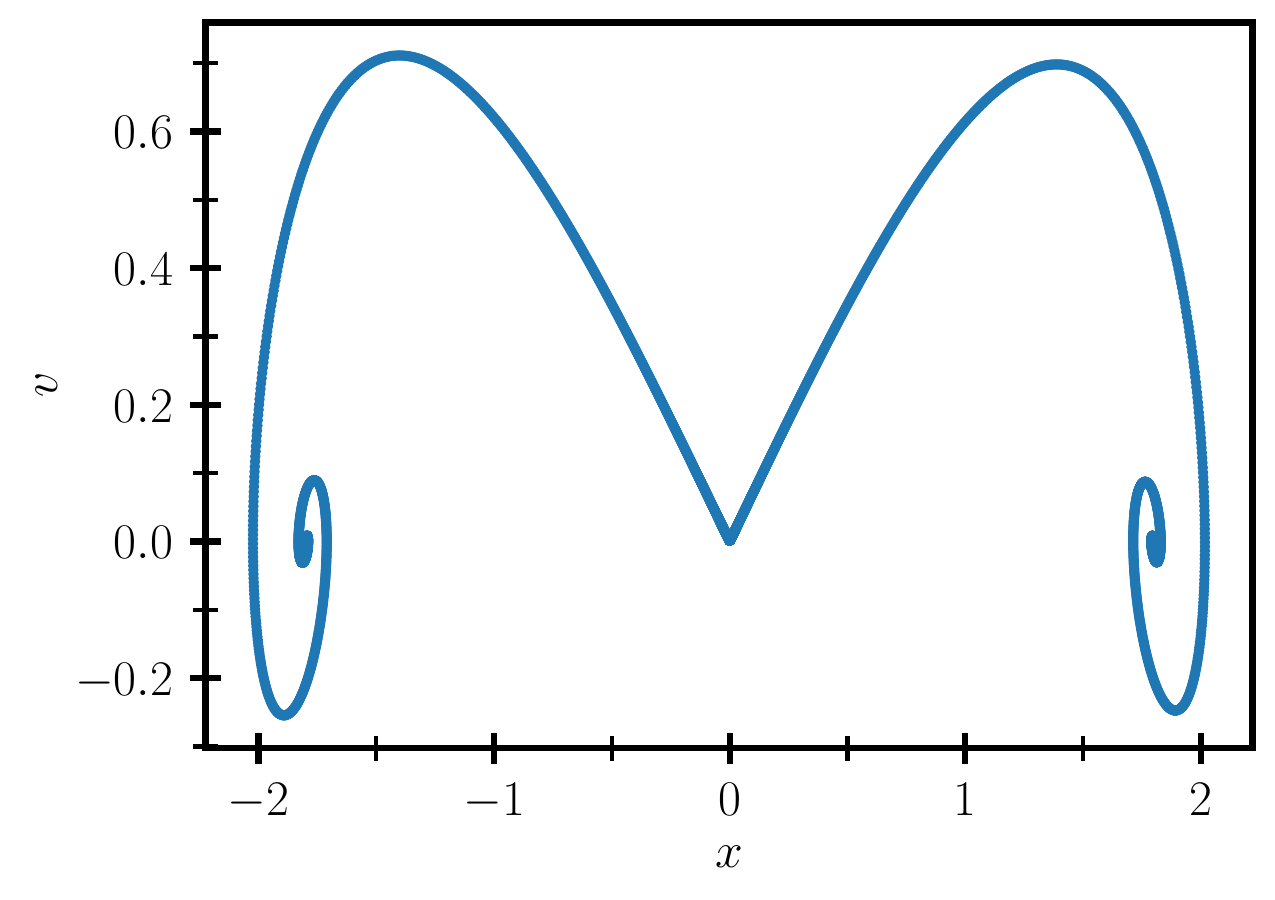}
    \caption{\rred{Underdamped instanton pathway in phase space for potential $V(x) = (x^2 - B^2)^2/B^4$ with $B=1.8$, $m=1$ and $\gamma=1$.}
    }
    \label{fig:Path_ud}
\end{figure}
In the main text, we focus on dynamics described by an overdamped Langevin equation, as is common in most condensed-phase environments, where inertial motion is rapidly quenched. However, some systems exhibit degrees of freedom with weak friction, for which the inertial motion becomes significant. In these cases, the underdamped Langevin equation must be employed
\begin{align}
    \label{equ:ud1}
    m \ddot{\mat{x}}(t) + \gamma\dot{\mat{x}}(t) = \mat{F}[\mat{x}(t)] +  \sqrt{2 \epsilon \gamma} \,\eta(t),
\end{align}
where for notational simplicity the mass $m$ and friction $\gamma$ are taken to be the same for all degrees of freedom. Following \Refx{Lehmann2003}, \eqn{equ:ud1} can be written in the extended space $\vec{\mat{x}} = (\mat{x},\mat{v})$
\begin{align}
    \label{equ:ud2}
    \dot{\vec{\mat{x}}} &= \vec{\mat{F}} + \sqrt{2\epsilon} \, \vec{\mathbf{\sigma}} \vec{\mat{\eta}} ,
\end{align}
where 
\begin{align}
    \vec{\mat{F}} = \begin{pmatrix} \mat{v}\\ \frac{1}{m} \mat{F} - \frac{\gamma}{m} \mat{v} \end{pmatrix} , \qquad 
    \vec{\mathbf{\sigma}} &= \sqrt{\frac{\gamma}{m^2}} \begin{pmatrix} \delta \, \mathbb{I}  & 0\\ 0 & \mathbb{I} \end{pmatrix} ,
\end{align} 
and $\vec{\mathbf{D}} = \vec{\mathbf{\sigma}}^\text{T} \vec{\mathbf{\sigma}}$. Here, an auxiliary noise scaled by $\delta$ is introduced to avoid a singular $\vec{\mathbf{D}}$ matrix. 
The regularization factor $\delta$ imposes the constraint $\dot{\mat{x}} = \mat{v}$ and can thus be viewed as a Lagrange multiplier, whose optimal value is zero. 
In practice, $\delta$ can simply be set to a constant that is small enough to tightly enforce the constraint. The instanton and the rate constant are then independent of the precise value of $\delta$, which is easily tested numerically.  Alternatively, a Legendre transform to a Hamiltonian formulation of the instanton dynamics could be employed \cite{Lehmann2003}.
The action for an underdamped system is given by
\begin{align}
    S &=  \int_{0}^{t_\text{f}} \vec{\mat{p}}^{\,\text{T}}\! \vec{\mathbf{D}} \vec{\mat{p}} \, \rmd t  ,
\end{align}
where $\vec{\mat{p}} = \vec{\mathbf{D}}^{-1} [\dot{\vec{\mat{x}}} - \vec{\mat{F}}]/2$ is the canonical momentum along the instanton in the extended space.

In Fig.~\ref{fig:Path_ud}, we show an underdamped instanton for a one-dimensional double well in phase space. It can be seen that the inertial motion leads to vortex-like curves within the potential wells, corresponding to the particle bouncing back and forth before overcoming the barrier. Like a person on a swing, the underdamped system can thus build up momentum to facilitate a more efficient transition, a mechanism that is lacking in overdamped descriptions due to the absence of inertial motion.

Given the similarity between \eqn{equ:ud1} and \eqn{equ:langevin}, the 
derivation of the underdamped instanton rate is analogous to the derivation for overdamped systems but in the extended $\vec{\mat{x}}$-space. The boundary conditions for the activation path are $\vec{\mat{x}}_\text{R} = (\mat{x}_\text{R},0)$ and $\vec{\mat{x}}^\ddagger = (\mat{x}^\ddagger,0)$. The resulting underdamped instanton rate constant is given by 
\begin{align}
    \label{equ:k_neqi_ud}
    k &\sim \frac{1}{4\pi \tau}
    \sqrt{\frac{B_t}{\tau \vec{p}_\text{f}^{\,2} \Sigma}}  \, \eu{-S_N/\epsilon},
\end{align}
where the fluctuation determinant $\Sigma = \det'[2\tau \vec{\nabla}^2_\text{s} S_N] (\text{det}[\vec{\mathbf{D}}])^{N}$ is obtained from the second-derivative matrix of the action in the extended space. The time-translation mode and the reaction coordinate along $\vec{\mat{p}}_\text{f}$ are again excluded from the determinant, and $B_t = \int \dot{\vec{\mat{x}}}^{\,2} \,\rmd t$ is the Jacobian for the transformation of the integral over the time-translation mode.  
\begin{figure}[t]
    \centering
    \includegraphics[width=0.48\textwidth]{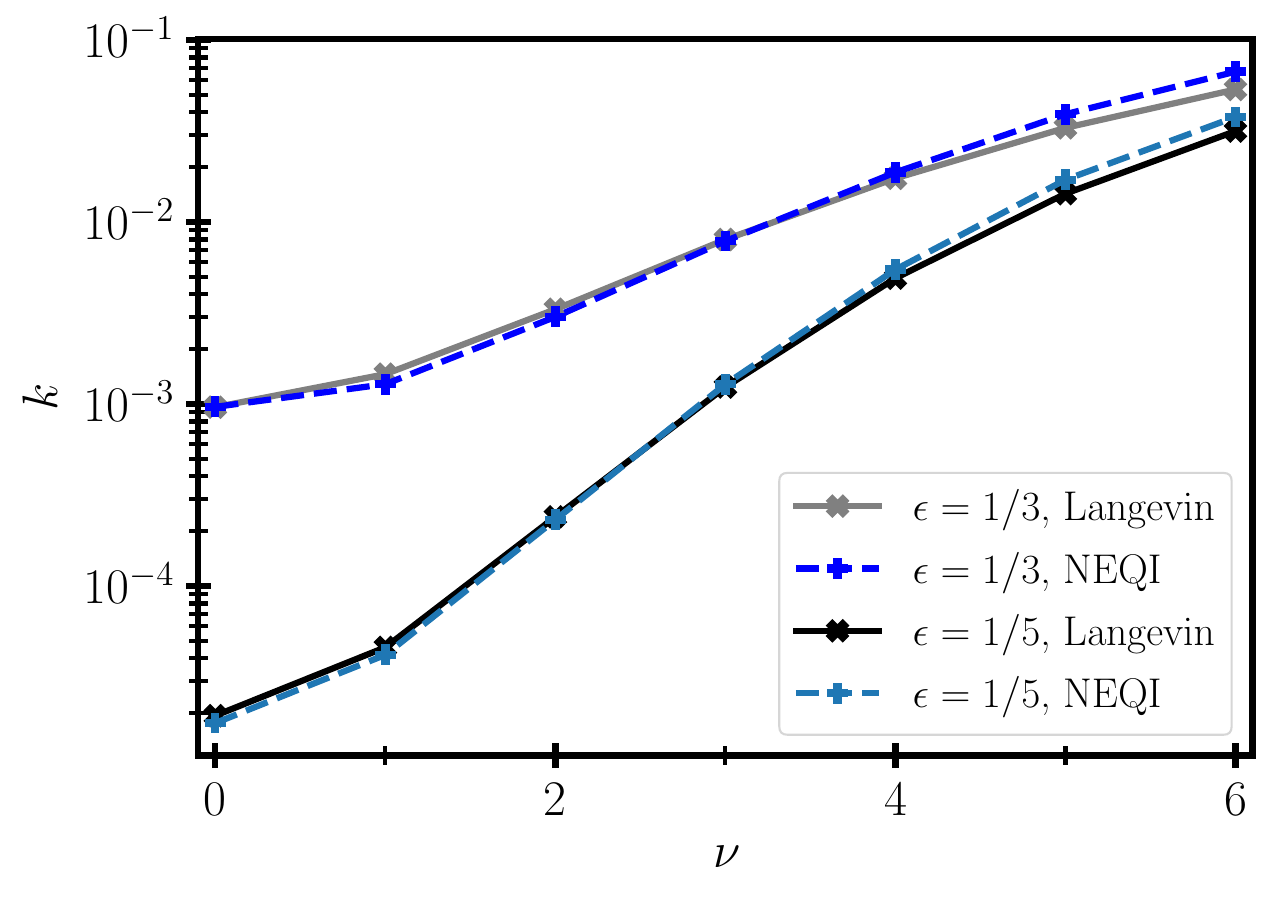}
    \caption{\rred{Rate constants for an underdamped system with $m=0.5$, $\gamma=2$ and the drift defined in \eqn{equ:force}. We compare results from the nonequilibrium instanton method (NEQI) to rates obtained from explicit simulations of the underdamped Langevin dynamics [\eqn{equ:ud1}] set up as described in the caption of Fig.~\ref{fig:mult_rates} for two different noise strengths $\epsilon$ over a range of driving strengths $\nu$.}  
    }
    \label{fig:Rates_ud}
\end{figure}
We illustrate the excellent agreement between the instanton rate constants and numerically exact results from direct integration of the underdamped Langevin dynamics at sufficiently weak noise in Fig.~\ref{fig:Rates_ud}.
However, it should be pointed out that, in analogy to \eqn{equ:ud_KLT}, the rate monotonically decreases with growing $\gamma$ and thus does not capture the energy-diffusion regime at very low friction \cite{Haenggi1990}. 

For systems at thermodynamic equilibrium, the instanton rate reduces to the underdamped Kramers--Langer rate
\begin{align}
    \label{equ:ud_KLT}
    k_\text{KLT} &= \frac{\lambda_\text{u}}{2\pi} \, \sqrt{\frac{\det(\nabla^2 V_\text{R})}{\lvert\det(\nabla^2 V^\ddagger)\rvert}}\, \eu{-\beta V^\ddagger} ,
\end{align}
where $\lambda_\text{u}$ is the sole positive eigenvalue of the Jacobian of the extended force, $\nabla \vec{\mat{F}}(\vec{\mat{x}}^\ddagger)$, evaluated at the TS\@.

}

\bibliography{references,other}
\end{document}